# Parametric study of upstream flame propagation in hydrogen-enriched premixed combustion: effects of swirl, geometry and premixedness

Ashoke De*

Department of Aerospace Engineering, Indian Institute of Technology Kanpur, Kanpur, 208016, India

Sumanta Acharya

Turbine Innovation and Energy Research Center, Louisiana State University, Baton Rouge, LA 70803, USA

**Abstract:** The effect of swirl, premixedness and geometry has been investigated for hydrogen enriched premixed flame using Large Eddy Simulation (LES) with a Thickened Flame (TF) model. Swirl strength has been varied to study the effects of swirl on flame behavior in a laboratory-scale premixed combustor operated under atmospheric conditions. In addition, the levels of premixedness and geometry have also been changed to study the role of these quantities on flame behavior. The turbulent flow field and the chemistry are coupled through TF model. In the LES-TF approach, the flame front is resolved on the computational grid through artificial thickening and the individual species transport equations are directly solved with the reaction rates specified using Arrhenius chemistry. Good agreement is found when comparing predictions with the published experimental data including the predicted RMS fluctuations. Also, the results show that higher swirl strength and increase in level of premixedness make the system more susceptible to upstream flame movement due to higher combustibility of hydrogen, which increases the reaction along the flame front, thereby raises temperature in the reaction zone and leads to combustion induced vortex breakdown (CIVB). Moreover, upstream flame movement is always observed at higher swirl strength irrespective of level of premixedness and burner geometry, whereas the premixed systems exhibit stable behavior while operating at low swirl.

*Corresponding Author: Tel.: +91-512-2597863 Fax: +91-512-2597561

E-mail address: ashoke@iitk.ac.in

## 1. INTRODUCTION

The current trend in the development of modern gas turbine combustor is to operate under lean conditions to reduce emissions. This can be best achieved by making the design more compact with low surface to volume ratio while maintaining the higher turbine inlet temperature. The compact design requires efficient mixing together with a compact premixed flame. A conceptual design may consist of a primary fuel injector nozzle, within which air passes through a swirler arrangement to properly mix the fuel with the incoming air. The turbulent swirling premixed fuel-air jet together with sudden expansion to the full combustor chamber provide an efficient way of improving mixing and help in stabilizing the flame. Hence due to the performance requirements, there is considerable interest in identifying the optimal swirl and geometric conditions to achieve specific practical goals.

One of the widely used techniques is jet-in-cross flow fuel injection systems due to its effectiveness and simplicity. The degree and rate of the mixing process is especially important in combustion applications. Boutazakhti et al. [1] studied the effect of jet mixing to improve the understanding of the relationship between mixing, chemical kinetics, and combustion efficiency for air jets in a hot reacting cross-flow. Other parameters such as the residence time and the equivalence ratio [2, 3] are equally important factors affecting combustion efficiency of air jets in a reacting cross-flow.

The other key factor in premixed combustion systems is Swirl. Swirl has great impact on the performance of gas turbine combustors including fuel-air pre-mixedness. Swirl is usually used to obtain high mixing rates as well as to stabilize the flames. Generally, the swirling motion at the inlet is generated using some kind of guide vanes, inlet tangential flow injections or by other means. When the strength of the swirl becomes high enough, it leads to formation of internal recirculation zone (IRZ), which is known as vortex breakdown phenomenon in fluid mechanics. Especially in Lean Premixed (LP) combustion, IRZ plays an important role by holding the hot combustion products and radicals as well as enhance the flame anchoring to the recirculation zone. However in the non-premixed combustion, IRZ usually enhances fuel/air mixing of fuel and intensity of combustion as well. Moreover, it has important influence on the flame shape, flame stability and heat release rate, as well as on emission. An extensive review on swirling flows can be found in [4-6]. Tangirala et al. [7] studied a non-premixed swirl burner where they reported that the mixing and flame stability can be improved with swirl upto a swirl number of about unity, beyond which a further increase in swirl reduces the turbulence level as well as the flame stability. Broda et al [8] and Seo [9] experimentally investigated the combustion dynamics in a lean-premixed swirl stabilized combustor. One of the key influences of the swirl on the flow field is the formation of the recirculation zone at the burner exit. As the swirl number exceeds a critical value, vortex breakdown takes place and leads to the formation of an internal recirculation zone [10]. This recirculation not only enhances fuel-air mixing, but also carries hot products back to the reactants and plays an important role in the flame holding. However, despite several years of research, the mechanisms of vortex breakdown are only partially understood [5, 11-12]. Recently, a number of studies have analyzed swirling flames using numerical approaches and experimental techniques as well [6].

In recent times, hydrogen enrichment has been found to be a promising approach of increasing the Lean Blow Out (LBO) limits of hydrocarbon fuels, thereby can provide stable combustion at lean mixture conditions [13]. Hence the problem of flame holding at LP conditions can be avoided by increasing the hydrogen proportion in the fuel. Increase in hydrogen proportion will assure better flame stability but, at the very same time, it will make the combustor more susceptible to flashback. Thus flashback becomes an inherent problem in this type of system due to addition of hydrogen, since hydrogen flame speeds are quite high [14]. It is an intrinsic behavior of premixed systems as the flame stabilizes at upstream of the combustion chamber. Flashback has been extensively investigated by several researchers [15-17] and they reported that the complexity of the topic becomes more substantial in swirling flows [18]. There are several explanations found in the literature [19-21], but the occurrence of flashback is unique to each system. Flashback may occur due to different reasons as: (a) Flashback by flame propagation, (b) Flashback in boundary layers, (c) Flashback due to turbulent flame speed, (d) Flashback due to combustion instabilities, and (d) Flashback due to vortex breakdown. Among all of these, flashback in strongly swirled flow mostly occurs either due to combustion instabilities or vortex breakdown. Although swirled burners being more sensitive to combustion instabilities, and may become prone to triggering the flashback in these types of combustors; however, the fluctuations required to cause such problems are beyond the acceptable noise levels in most combustion systems.

Vortex breakdown in swirled burner occurs when the azimuthal velocity becomes larger than the axial velocity. Moreover, this vortex breakdown is often accompanied by a large recirculation zone with high reverse flow velocities in combustions systems, and the reverse velocities can promote upstream flame propagation and flashback is the consequence. The formation of vortex breakdown strongly depends on the geometry, and if the swirl number exceeds a critical value the recirculation zone is able to extend itself throughout the entire mixing section. In isothermal flows, this effect can be prevented by selecting the swirl number; however, the chemical reaction can nevertheless lead to a breakdown of the flow, combined with upstream flame propagation. This mechanism is called as combustion-induced vortex breakdown (CIVB) as reported by Fritz et al. [19], Sommerer et al. [22]. Normally, swirled burner and swirl stabilized flame without center-body are more susceptible to this kind of problem. Recent studies by Nauert et al. [23], Kiesewetter et al. [24], Kröner et al. [20, 25], Knole et al. [26], Konle and Sattelmayer [27], Voigt et al. [28], Hegger et al. [29], Tangermann et al. [30] also reported this CIVB driven flashback in swirled burner.

Upstream flame propagation into the pre-mixer section leads to thermal overload and destruction of the hardware therefore it must be avoided at all load conditions [24]. This can be prevented by using specially designed flame holders or by injecting syngas in a separate non-premixed arrangement. However, in transitioning from natural gas as the fuel of choice to syngas, it is desirable to keep hardware changes to a minimum, given the extensive body of knowledge with current natural gas related hardware. Several studies have investigated premixed flames of $H_2$-hydrocarbon fuel mixtures in swirled burner. Schefer [31] and Schefer et al. [32] studied the effects of hydrogen injection in methane/air flames in a completely premixed mode combustor at atmospheric pressure and swirling conditions. They reported that the addition of $H_2$ to methane ($CH_4$) fuel decreases the adiabatic flame temperature at LBO and, hence, decreases the CO emissions (without effecting the NOx emissions). The $H_2$-enriched $CH_4$ flame produces shorter flames with more intense reaction zones. Gupta et al. [33-34] investigated the effects of swirl on combustion characteristics of premixed flames. Kröner et al. [20, 25] also reported flashback due to combustion induced vortex breakdown (CIVB) in swirling flows for different $CH_4+H_2$ mixtures. They reported that the CIVB happened to be the prevailing mechanism in a swirled burner without center-body. Strakey et al. [35] in their study investigated the effects of hydrogen addition on lean extinction in a swirl stabilized combustor. They reported the lean blowout limits for methane/hydrogen mixtures at pressures ranging from 1 to 8 atmospheres. More recently, Kim et al. [36-37] reported the effects of $H_2$ addition in a confined and un-confined swirl burner operating at lean conditions through PIV diagnostics. They clearly showed the impact of hydrogen on flame structure and flow field including the pollutant emissions. They also reported the effects of swirl intensity on hydrogen enriched flame how that alters the flow field. Tuncer et al. [38] also investigated dynamics, NOx and flashback characteristics in a confined premixed hydrogen-enriched methane flames for a laboratory scale swirled combustor. Bellester et al. [39] also carried out chemuluminescence measurements to study premixed natural gas flames with hydrogen blending for a swirl-stabilized combustor.

In this work, a Thickened Flame (TF) model [40] is used where the flame is artificially thickened to resolve it on computational grid points where reaction rates from kinetic models are specified using reduced mechanisms. The influence of turbulence is represented by a parameterized efficiency function. A key advantage of the TF model is that it directly solves the species transport equations and uses the Arrhenius formulation for the evaluation of the reaction rates.

In turbulent premixed combustion, a popular approach is to rely on the flamelet concept, which essentially assumes the reaction layer thickness to be smaller than the smallest turbulence scales. The two most popular model based on this concept are the flame surface density model (FSD) [41] and the G-equation model [42-43]. It has been reported that the FSD model is not adequate beyond the corrugated flamelet regime [44-45], while the G-equation approach depends on a calculated signed-distance function that represents an inherent drawback of this method.

Another family of models relies on the probability density function (PDF) approach [43, 46], which directly considers the probability distribution of the relevant quantities in a turbulent reacting flow. Moreover, it can be applied to non-premixed, premixed, and partially premixed flames without having much difficulty. Usually, there are two ways which are mainly used to calculate the pdf: one is *presumed pdf approach*, and other is *pdf transport balance equation approach*. The *presumed pdf approach*, which essentially assumes the shape of the probability function P, is relatively simpler to use, however, has severe limitations in the context of applicability. On the other hand, the *pdf transport balance equation approach* solves a transport equation for pdf function, which is applicable for multi species, mass-weighted probability density function. This method has considerable advantage over any other turbulent combustion model due to its inherent capability of handling any complex reaction mechanism. However, the major drawback of transport pdf approach is its high dimensionality, which essentially makes the implementation of this approach to different numerical techniques, like FVM or FEM, limited, since their memory requirements increase almost exponentially with dimensionality. Usually, Monte-Carlo algorithms, which reduce the memory requirements, are used by Pope [46]. Moreover, a large number of particles need to be present in each grid cell to reduce the statistical error; however this makes it a very time consuming process. So far, the transport equation method has been only applied to relatively simple situations.

The major advantage associated with this TF model is the ability to capture the complex swirl stabilized flame behavior which is often found in a gas turbine combustor. Since this type of geometry with the premixing section does not guarantee a perfectly premixed gas at the dump plane, the fully premixed assumption in the numerical model is not valid any more. The present TF model is capable of taking care of this type of partially premixed gas since we solve for the individual species transport equations and the reaction rates are specified using Arrhenius expressions.

The configuration of interest in the present work is an unconfined swirl-stabilized flame. In this investigation, we investigate upstream flame propagation behavior of hydrogen enriched premixed swirl-stabilized flames. The goal of this study is to analyze the flow and combustion physics in hydrogen enriched premixed flame and, in particular, to explore how swirl, premixednes and geometry play a role in the flame behavior.

## 2. FLOW CONFIGURATION

The configuration considered here is an unconfined swirl burner as shown in Fig. 1 [6]. The 45° swirl vane is fitted with a solid center body which also acts as a fuel injector [6]. This center body extends beyond the swirl vane and is flush with the dump plane of the combustor. The diameter of the center body is 12.7mm (0.5 inch) and the outer diameter (O.D.) of the swirler is 34.9 mm (1.375 inch). Methane and hydrogen gas is injected radially from the center body through eight holes immediately downstream of the swirler vane. The fuel/air mixer is assumed to be perfectly premixed at the dump plane and the equivalence ratio is calculated to be $\varphi$=0.7. The investigation is carried out for Reynolds number, Re=13339 (based on inlet bulk velocity and hydraulic diameter) at atmospheric pressure and temperature with 30%$H_2$ mixture. The swirl number, defined as the ratio of the axial flux of the tangential momentum to the product of axial momentum flux and a characteristic radius, used for this investigation are S=0.38, 0.82 and 1.76.

## 3. NUMERICAL DETAILS

### 3.1 Governing equations and flow modeling using LES

The filtered governing equations for the conservation of mass, momentum, energy and species transport are given as:

Continuity equation:

$$\frac{\partial}{\partial t}(\bar{\rho}) + \frac{\partial}{\partial x_i}(\bar{\rho}\bar{u}_i) = 0 \tag{1}$$

Momentum equation:

$$\frac{\partial}{\partial t}(\bar{\rho}\bar{u}_i) + \frac{\partial}{\partial x_j}(\bar{\rho}\bar{u}_i\bar{u}_j) = -\frac{\partial}{\partial x_i}(\bar{p}) + \frac{\partial}{\partial x_j}\left((\mu + \mu_t)\frac{\partial \bar{u}_i}{\partial x_j}\right) \tag{2}$$

Energy equation:

$$\frac{\partial}{\partial t}(\bar{\rho}\bar{E}) + \frac{\partial}{\partial x_i}(\bar{\rho}\bar{u}_i\bar{E}) = -\frac{\partial}{\partial x_j}\left(\bar{u}_j\left(-\bar{p}I + \mu\frac{\partial \bar{u}_i}{\partial x_j}\right)\right) + \frac{\partial}{\partial x_i}\left(\left(k + \frac{\mu_t C_p}{\mathrm{Pr}_t}\right)\frac{\partial \bar{T}}{\partial x_i}\right)$$
$$+ \frac{\partial}{\partial x_i}\left(\bar{\rho}\sum_{s=1}^{N} h_s\left(D + \frac{\mu_t}{Sc_t}\right)\frac{\partial \bar{Y}_s}{\partial x_i}\right) \tag{3}$$

Species transport equation:

$$\frac{\partial}{\partial t}(\bar{\rho}\bar{Y}_i) + \frac{\partial}{\partial x_j}(\bar{u}_j\bar{\rho}\bar{Y}_i) = \frac{\partial}{\partial x_j}\left(\left(D + \frac{\mu_t}{Sc_t}\right)\frac{\partial \bar{Y}_i}{\partial x_j}\right) + \dot{\bar{\omega}}_i \tag{4}$$

where $\rho$ is the density, $u_i$ is the velocity vector, $p$ is the pressure, $E = e + u^2_i /2$ the total energy, where $e = h - p/\rho$ is the internal energy and $h$ is enthalpy, $\mu$ is viscosity, $k$ is thermal conductivity, $D$ is molecular diffusivity, $\mu_t$ is turbulent eddy viscosity, $Sc_t$ is turbulent Schmidt number, $\mathrm{Pr}_t$ is turbulent Prandtl number, $h_f^0$ is enthalpy of formation, and $\dot{\omega}$ is reaction rate.

To model the turbulent eddy viscosity, LES is used so that the energetic larger-scale motions are resolved, and only the small scale fluctuations are modeled. The sub-grid stress modeling is done using a dynamic Smagorinsky model where the unresolved stresses are related to the resolved velocity field through a gradient approximation:

$$\overline{u_i u_j} - \overline{u_i}\,\overline{u_j} = -2\nu_t \overline{S}_{ij} \tag{5}$$

Where
$$\nu_t = C_s{}^2 (\Delta)^2 \left| \overline{S} \right| \tag{6}$$

$$\overline{S}_{ik} = \frac{1}{2}\left( \frac{\partial \overline{u_i}}{\partial x_k} + \frac{\partial \overline{u_k}}{\partial x_i} \right) \tag{7}$$

$$\left| \overline{S} \right| = \sqrt{2 \overline{S}_{ik}\,\overline{S}_{ik}} \tag{8}$$

and S is the mean rate of strain. The coefficient $C_s$ is evaluated dynamically [47-48] and locally-averaged.

## 3.2 Combustion modeling

Modeling of flame-turbulence interaction in premixed flames requires tracking of the thin flame front on the computational grid. In the present work, we used TF modeling technique, where the flame front is artificially thickened to resolve on computational grid. Corrections are made to ensure that the flame is propagating at the same speed as the un-thickened flame [40, 49]. The key benefit of this approach, as noted earlier, rests in the ability to computationally resolve the reaction regions and the chemistry in these regions. More details on this approach are described in the following sections.

**Thickened-Flame (TF) Modeling Approach with LES:**

Butler and O'Rourke [50] were the first to propose the idea of capturing a propagating premixed flame on a coarser grid. The basic idea with this approach is that the flame is artificially thickened to include several computational cells and by adjusting the diffusivity to maintain the same laminar flame speed $s_L^0$. It is well known from the simple theories of laminar premixed flames [51-52] that the flame speed and flame thickness can be related through the following relationship

$$s_L^0 \propto \sqrt{D\overline{B}}, \delta_L^0 \propto \frac{D}{s_L^0} = \sqrt{\frac{D}{\overline{B}}} \tag{9}$$

where D is the molecular diffusivity and $\overline{B}$ is the mean reaction rate. When the flame thickness is increased by a factor F, the molecular diffusivity and reaction rate are modified accordingly (FD and $\overline{B}$/F) to maintain the same flame speed. The major advantages associated with thickened flame modeling are: (i) the thickened flame front is resolved on LES mesh which is usually larger than typical premixed flame thickness (around 0.1-1 mm), (ii) quenching and ignition events can be simulated, (iii) chemical reaction rates are calculated exactly like in a DNS calculation without any *ad-hoc* sub models, so it can theoretically be extended to incorporate with multi-step chemistry [40].

In LES framework, the spatially filtered species transport equation is given in Equation (4), where the terms on the right hand side are the filtered diffusion flux plus the unresolved transport, and the filtered reaction rate respectively. In general, the unresolved term is modeled with a gradient diffusion assumption by which the laminar diffusivity is augmented by the turbulent eddy diffusivity. However, in the TF model, the "thickening" procedure multiplies the diffusivity term by a factor F which has the effect of augmenting the diffusivity. Therefore, the gradient approximation for the unresolved fluxes is not explicitly used in the closed species transcription equations. The corresponding filtered species transport equation in the thickened-flame model becomes

$$\frac{\partial \overline{\rho Y_i}}{\partial t} + \frac{\partial}{\partial x_j}(\overline{\rho Y_i u_j}) = \frac{\partial}{\partial x_j}\left(\overline{\rho}FD_i\frac{\partial \overline{Y_i}}{\partial x_j}\right) + \frac{\overline{\dot{\omega}_i}}{F} \tag{10}$$

Although the filtered thickened flame approach looks promising, a number of key issues need to be addressed. The thickening of the flame by a factor of F modifies the interaction between turbulence and chemistry, represented by the Damköhler number, Da, which is a ratio of the turbulent ($\tau_t$) and chemical ($\tau_c$) time scales. Da, is decreased by a factor F and becomes Da/F, where

$$Da = \frac{\tau_t}{\tau_c} = \frac{l_t s_L^0}{u' \delta_L^0} \tag{11}$$

As the *Da* is decreased, the thickened flame becomes less sensitive to turbulent motions. Therefore, the sub-grid scale effects have been incorporated into the thickened flame model, and parameterized using an efficiency function E derived from DNS results [40]. Using the efficiency function, the final form of species transport equation becomes

$$\frac{\partial \overline{\rho Y_i}}{\partial t} + \frac{\partial}{\partial x_j}(\overline{\rho Y_i u_j}) = \frac{\partial}{\partial x_j}\left(\overline{\rho}EFD_i\frac{\partial \overline{Y_i}}{\partial x_j}\right) + E\frac{\overline{\dot{\omega}_i}}{F} \tag{12}$$

where the modified diffusivity ED, before multiplication by F to thicken the flame front, may be decomposed as ED=D(E-1)+D and corresponds to the sum of molecular diffusivity, D, and a turbulent sub-grid scale diffusivity, (E-1)D. in fact, (E-1)D can be regarded as a turbulent diffusivity used to close the unresolved scalar transport term in the filtered equation.

The central ingredient of the TF model is the sub-grid scale wrinkling function E, which is defined by introducing a dimensionless wrinkling factor $\Xi$. The factor $\Xi$ is the ratio of flame surface to its projection in the direction of propagation. The efficiency function, E, is written as a function of the local filter size ($\Delta_e$), local sub-grid scale turbulent velocity ($u'_{\Delta_e}$), laminar flame speed ($s_L^0$), and the thickness of the laminar and the artificially thickened flame ($\delta_L^0, \delta_L^1$). Colin et al. [40] proposed the following expressions for modeling the efficiency function.

$$\Xi = 1 + \beta \frac{u'_{\Delta_e}}{s_L^0} \Gamma\left(\frac{\Delta_e}{\delta_L^0}, \frac{u'_{\Delta_e}}{s_L^0}\right) \tag{13}$$

$$\beta = \frac{2\ln 2}{3c_{ms}\left(\mathrm{Re}_t^{1/2} - 1\right)}, c_{ms} = 0.28, \mathrm{Re}_t = u'l_t / \nu \tag{14}$$

where $\mathrm{Re}_t$ is the turbulent Reynolds number. The local filter size $\Delta_e$ is related with laminar flame thickness as

$$\Delta_e = \delta_L^1 = F\delta_L^0 \tag{15}$$

The function $\Gamma$ represents the integration of the effective strain rate induced by all scales affected due to artificial thickening, $\Gamma$ is estimated as

$$\Gamma\left(\frac{\Delta_e}{\delta_L^0}, \frac{u'_{\Delta_e}}{s_L^0}\right) = 0.75 \exp\left[-1.2\left(\frac{u'_{\Delta_e}}{s_L^0}\right)^{-0.3}\right]\left(\frac{\Delta_e}{\delta_L^0}\right)^{2/3} \tag{16}$$

Finally, the efficiency function takes the following form as defined by the ratio between the wrinkling factor, $\Xi$, of laminar flame ($\delta_L = \delta_L^0$) to thickened flame ($\delta_L = \delta_L^1$).

$$E = \frac{\Xi\big|_{\delta_L=\delta_L^0}}{\Xi\big|_{\delta_L=\delta_L^1}} \geq 1 \qquad (17)$$

where the sub-grid scale turbulent velocity is evaluated as $u'_{\Delta_e} = 2\Delta_x^3 \left| \nabla^2 \left( \nabla \times \bar{u} \right) \right|$, and $\Delta_x$ is the grid size. This formulation for sub-grid scale velocity estimation is free from dilatation. Usually, $\Delta_e$ differs from $\Delta_x$, and it has been suggested that values for $\Delta_e$ be at least $10\Delta_x$ [40].

There are different versions of TF model available in literature depending on the calculation of E such as: Power-law flame wrinkling model [53-54], Dynamically modified TF model [55-56]. In one of the previous studies by the authors reported that these different versions of TF models do not exhibit substantial differences in flow filed and temperature predictions [57]. That's why we use the original formulation of TF model for the present study [40]. More detailed description of E can found in other literature [58].

### 3.3 Chemistry model

As all the species are explicitly resolved on the computational grid, the TF model is best suited to resolve major species. Intermediate radicals with very short time scales can not be resolved. To this end, only simple global chemistry has been used with the thickened flame model.

For $CH_4$ combustion, a two-step chemistry, which includes six species ($CH_4$, $O_2$, $H_2O$, $CO_2$, $CO$ and $N_2$) is used and given by the following equation set.

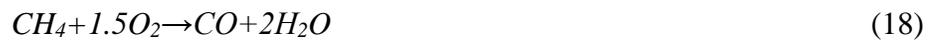

$$CH_4 + 1.5O_2 \rightarrow CO + 2H_2O \qquad (18)$$

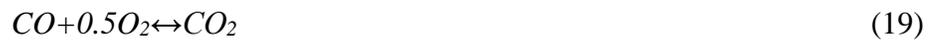

$$CO + 0.5O_2 \leftrightarrow CO_2 \qquad (19)$$

To incorporate $H_2$ reaction in addition to the above $CH_4$ chemistry, the following 1-step Marinov mechanism is employed.

$$H_2+0.5O_2 \rightarrow H_2O \tag{20}$$

The corresponding reaction rate expressions are given by:

$$q_1=A_1 exp(-E^1_a/RT)[CH_4]^{a1}[O_2]^{b1} \tag{21}$$

$$q_2(f)=A_2 exp(-E^2_a/RT)[CO][O_2]^{b2} \tag{22}$$

$$q_2(b)=A_2 exp(-E^2_a/RT)[CO_2] \tag{23}$$

$$q_3=A_3 exp(-E^3_a/RT)\,[H_2][O_2]^{b3} \tag{24}$$

where the activation energy $E^1_a$ =34500 cal/mol, $E^2_a$ =12000 cal/mol, a1=0.9, b1=1.1, b2=0.5, and $A_1$ and $A_2$ are 2.e+15 and 1.e+9, as given by Selle et al. [59], and $E^3_a$ =35002 cal/mol, b3=0.5, $A_3$=1.8e+16 (SI units) as given in the DOE report [60]. The first and third reactions (Eqs. 18 & 20) are irreversible, while the second reaction (Eq. 19) is reversible and leads to an equilibrium between CO and $CO_2$ in the burnt gases. Hence the expressions (Eqs. 21 & 24) represent the reaction rates for the irreversible reactions (Eqs. 18 & 20) and the expressions (Eqs. 22 & 23) represent the forward and backward reaction rates for the reversible reaction (Eq. 19). Properties including density of mixtures are calculated using CHEMKIN-II [61] and TRANFIT [62] depending on the local temperature and the composition of the mixtures at 1 atm.

## 3.4 Solution procedure

In the present study, a parallel multi-block compressible flow code for an arbitrary number of reacting species, in generalized curvilinear coordinates, is used. Chemical mechanisms and thermodynamic property information of individual species are input in standard Chemkin format. Species equations along with momentum and energy equation are solved implicitly in a fully coupled fashion using a low Mach number preconditioning technique, which is used to effectively rescale the acoustics scale to match that of convective scales [63]. An Euler differencing for the pseudo time derivative and second order backward 3-point differencing for physical time derivatives are used. A second order low diffusion flux-splitting algorithm is used for convective terms [64]. However, the viscous terms are discretized using second order central differences. An incomplete Lower-Upper (ILU) matrix decomposition solver is used. Domain decomposition and load balancing are accomplished using a family of programs for partitioning unstructured graphs and hypergraphs and computing fill-reducing orderings of sparse matrices, METIS. The message communication in distributed computing environments is achieved using Message Passing Interface, MPI. The multi-block structured curvilinear grids presented in this paper are generated using commercial grid generation software GridPro$^{TM}$.

### 3.5 Computational domain and boundary conditions

As shown in Fig. 1, the configuration of interest in the present work is an unconfined swirled burner [6]. The computational domain extends 20D downstream of the dump plane (fuel-air nozzle exit), 13D upstream of the dump plane (location of the swirl vane exit) and 6D in the radial direction. Here, D is the center-body diameter. The finer mesh consists of 320x208x48 grid points downstream of the dump plane plus (98x32x48) + (114x22x48) grid points upstream, and contains approximately 3.94M grid points [6]. The grid resolution in the computational domain with 3.94M grid points is given as: (a) along axial direction $\Delta x/2D=0.031$ is maintained from inlet to dump plane, and then 0.03125 rest of the whole domain starting from dump-plane to outlet, (b) along the radial direction $\Delta r/2D=0.014$-$0.0145$ is maintained starting from centerline to lateral boundary, (c) along azimuthal direction: behind center-body $\Delta\theta/2D=0.011$-$0.0327$ (up to r/2D=0.25), in the annular shear layer $\Delta\theta/2D=0.0327$-$0.089$ (r/2D=0.25-0.6875), and finally Δθ/2D=0.089-0.393 (r/2D=0.6875-3.0) where D=center-body diameter. More details on computation grid can be found in the literature [6].

The inflow boundary condition is assigned at the experimental location immediately downstream of the swirler blades. The mean axial velocity distribution is specified as a one-seventh power law profile to represent the fully developed turbulent pipe flow, with superimposed fluctuations at 10% intensity levels (generated using Gaussian distribution). A constant tangential velocity component is specified as determined from the swirl vane angle. Convective boundary conditions [65] are prescribed at the outflow boundary, and zero-gradient boundary conditions are applied on the lateral boundary. The time step used for the computation is dt=1.0e-3 where the flow through time is ~3.2sec. The fuel injection point is shown in Fig. 1, which represents a jet-in-cross flow type configuration.

## 4. RESULTS AND DISCUSSION

We will first report the non-reacting LES calculations to ensure that the grid and boundary conditions are properly chosen, and to assess the cold-flow flow characteristics. This will be followed by a discussion of the hydrogen enriched flow calculations where we will examine upstream flame propagation behavior and analyze how swirl, geometry and premixedness influence this behavior.

### 4.1 Non-reacting flow results

Figure 2 shows the radial distribution of axial and tangential mean velocity profiles, axial and tangential velocity fluctuations at different axial locations for Re=13339. In general, LES and the experimental data for radial distribution of the axial and tangential mean velocity profiles, and the axial and tangential fluctuations at different axial locations are found to be in good agreement. It is observed that the shape, size, and the intensity of the recirculation zone (region of negative axial velocities at the center) are well predicted along with the overall spreading of the turbulent swirling jet. Also, the RMS fluctuations of the axial and tangential velocities are in good agreement with the experimental data. The peak in the axial velocity fluctuations is observed to be in the shear layer and between the location of the peak velocity and the recirculation bubble. In this region, the steepest velocity gradient $\partial U_i/\partial x_j$ is obtained and promotes the production of the peak kinetic energy. The tangential velocity fluctuations show a flatter profile than the axial velocity fluctuations and their peaks are shifted radially inwards as for the mean tangential velocities. Since the fine mesh (3.94M grid points) results are found to be in better agreement with the experimental data, this fine mesh is chosen for reacting flow calculations. More detailed discussion on non-reacting flow results can be found in the literature by De et al. [6].

**4.2 Reacting flow results**

In this section we will present all the reacting flow calculations for Re=13339 with 30%H2 mixture.

### 4.2.1 Effects of swirl

The flow configuration used here is the jet-in-cross flow fuel injection system. Means, the fuel is injected right after the step (Fig. 1) and the inlet swirled air is mixed with the injected fuel to have a premixed mixture at dump plane.

In order to study the effects of swirl, different swirl strength (S=0.38, S=0.82, S=1.76) for Re=13339 with 30%$H_2$ has been considered herein. Usually the swirl enhances fuel-air mixing to achieve premixed mixture at the dump plane of combustor chamber. Hence increasing swirl strength increases the mixing, at the same time increases the turbulence level as well. The effect of different swirl strength is clearly observed in axial velocity profiles. As observed the magnitude of the peak axial velocity increases by ~8% (S=0.38-S=0.82) and ~30% (S=0.38-S=1.76) with the increase in swirl strength. The reason being the higher swirl produces more turbulence, which, in turn increases the heat release due to higher turbulence-chemistry interaction; hence flame temperature increases and peak density decreases by ~11% (S=0.38-S=0.82) and ~28% (S=0.38-S=1.76). Thus flow accelerates more along the flame front and exhibit higher magnitude. However, at a fixed hydrogen mixture (30%$H_2$), the increase in swirl strength increases the size of the recirculation bubble (shown in Fig. 3(a)), thus the overall recirculation flow increases. As observed the width of the recirculation zone is also broadened due to higher centrifugal force at higher swirl strength (Fig. 3(a)). Similar trend is also reported in the literature by Kim et al. [36-37].

The radial distributions of mean axial velocity and axial velocity fluctuations are shown in Fig. 3(b) for S=0.82. The overall agreement of the predictions with the data is found to be quite reasonable, considering the complexity of the physical processes and the configuration. With increasing axial distance the magnitude of the peak velocity decreases and the location of the peak is moved further outwards radially. While the general agreement between the data and predictions are satisfactory, and the LES results show the right qualitative features and the peak magnitudes, there are intrinsic differences between the predictions and data. The behavior of predicted RMS fluctuations can be associated with the higher temperature in the product region (Fig. 4). This corresponds to the burnt and un-burnt regions in the inner part of the shear layer and associated with the high velocity gradients where the turbulence production due to the mean velocity gradient is the highest. The lower magnitude is located in the burnt region of the shear layer downstream of the center body where the temperatures are higher. The high temperatures cause the viscosity value to go up, and this reduces the magnitude of the peak stress component.

The higher magnitude is observed in the un-burnt regions of the shear layer where the temperatures are relatively on the lower side which reduces the viscosity value and in turn exhibits higher fluctuating components. Similar trends for RMS fluctuations have also reported by other researchers reporting calculations [66].

Figure 4 shows time series temperature plots for different S, which clearly shows how the flame starts propagating upstream and then stabilizes in the mixing tube, especially at higher swirl strength, i.e. S=0.82 and S=1.76. This can be related to the high level of turbulence generated at higher swirl in addition to the higher low velocity region along the wall of the center-body. Higher turbulence-chemistry interaction increases heat release production, especially in $H_2$ enriched mixture, and promotes this upstream movement. For S=0.38, the flame front also moves little bit upstream of the dump plane but does not move further upstream as observed in Fig. 4. Upstream flame propagation, usually characterized as flame flashback, occurs when the burning velocity exceeds the local flow velocity. In the present case, the upstream propagation is initiated due to Combustion Induced Vortex Breakdown (CIVB) and then accompanied with the favorable condition, due, in part, low velocity in B/L and higher local burning velocity. As the flame starts moving upstream, especially for S=0.82 and S=1.76, it encounters fuel rich region due to the fuel injection point, and that induces higher flame propagation velocity in presence of highly combustible $H_2$. More often, the CIVB refers to the vortex break down due to chemical reaction which becomes more intensive in presence of $H_2$ and turbulence, enhanced at higher swirl strength as well. Additionally, this vortex breakdown is often accompanied by a large recirculation zone with high reverse flow velocities in combustion systems, and the reverse velocities can promote this upstream flame propagation (Fig. 4).

Usually, the recirculation bubbles bring the heat and reactive species back to the flame tip and provides flame-holding at the dump plane. However, the recirculation zone gives rise to the azimuthal vorticity component in this region and that produces positive or negative induced velocity following the Bio-Savart law, as given by Eq. 27 [24, 26] depending on the sign of azimuthal vorticity. The position and motion of this recirculation bubble largely depends on the balance maintained between the induced velocity and irrotational axial velocity. Slight changes in the flow field can alter this balance and result in the upstream movement of the recirculation bubble. This unsteady motion of recirculation bubble actually initiates the upstream flame propagation in the mixing tube, and thereafter it is accompanied by the higher induced negative velocity due to production of negative azimuthal vorticity. As observed for S=0.38 in Fig. 5, the flame front moves little bit upstream into the mixing tube and tends to interact with the fuel injection system (Fig. 5) and form a small recirculation bubble ahead of flame front at t=9.2s (Fig. 5). Since the further upstream movement of the flame front is annihilated due to unfavorable condition, the flame front is actually pushed back by the incoming flow (t=11.2s & t=17.6s in Fig. 5) and the formed small recirculation bubble (t=9.2s & t=14.2 s in Fig. 5) also disappears at t=11.2s & t=17.6s (Fig. 5). This back-n-forth interaction of flame front and incoming flow maintains a balance between flow field and flame front without occurrence of upstream flame propagation at lower swirl strength. That's why the upstream flame propagation in this case is not observed. However, in the case of higher swirl strength S=0.82 & S=1.76, the upstream flame movement is also initiated due to unsteady motion of recirculation bubble, formed due to vortex breakdown. Figs. 6 and 7 depict the time instant snap shots of stream lines superimposed with temperature contours during upstream flame propagation for S=0.82 and S=1.76, respectively. As observed in Fig. 6 for S=0.82, the flame front is nicely stable close to the dump plane at t=9s and there is no interaction between flame front and jet-in-cross flow fuel injection. Once the flame front moves further upstream at t=12.2s due to upstream movement of stagnation point (formed due to vortex breakdown), it starts interacting with the fuel injection system due to increase in pressure ahead of flame front. The pressure difference between upstream and downstream of flame front causes the formation of a small recirculation bubble ahead of flame front and also forms a local stagnation point ahead of flame tip, which becomes more prominent later time instant at t=14.2s (Fig. 6). The formation of this small recirculation bubble ahead of the flame tip gives rise to the generation of greater negative azimuthal vorticity, which in turn, produces higher induced negative velocity and the flame tip movement to further upstream becomes completely uncontrollable. That's why this case exhibits upstream movement while S=0.38 case does not show such behavior (Fig.6).

In order to further illustrate the pressure jump, which causes to form recirculation bubble ahead of flame tip and giving rise to the negative induced velocity, across the flame front in the axial and radial direction, the simplified radial momentum equation is used : $\frac{\partial p}{\partial r} = \frac{\rho \overline{w}^2}{r}$, where ρ is the density, $\overline{w}$ is the tangential velocity, and r is the radial coordinate [29]. Assuming p=p(x,r), radial pressure gradient can be related as: $\frac{dp}{dr}\bigg|_{r_i}^{r_o} = \int_{r_i}^{r_o} \frac{\rho \overline{w}^2}{r} dr$, where $r_i$ is the radius at the center body wall and $r_o$ is the radius at the outer wall. Due to density difference between outer and inner wall across the flame front in the mixing tube, the radial pressure difference is always dp>0. Moreover, this dp increases with the increase in swirl strength (S) since higher S imposes higher tangential velocity component ($\overline{w}$). To maintain conservation of momentum, the axial pressure gradient also changes. However, the axial pressure gradient along the inner wall is larger than that along the outer wall as shown by the following relationship:

$$\left.\frac{dp}{dx}\right|_{r_o} - \left.\frac{dp}{dx}\right|_{r_i} = \int_{r_i}^{r_o} \frac{\partial}{\partial x}\left(\frac{\rho \overline{w}^2}{r}\right)dr => \left.\frac{dp}{dx}\right|_{r_i} = \left.\frac{dp}{dx}\right|_{r_o} - \int_{r_i}^{r_o}\left(\overline{w}^2 \frac{\partial \rho}{\partial x} + 2\rho\overline{w}\frac{\partial \overline{w}}{\partial x}\right)dr \quad (25)$$

The above expression clearly reveals that the axial pressure gradient along the inner wall is larger than that at outer wall due to decrease in density in the axial direction ($\frac{\partial \rho}{\partial x} < 0$) and decrease of tangential velocity in the axial direction as well. In the present scenario, higher swirl intensity contributes to the higher decrement in tangential velocity in axial direction, and consequently dp/dx always becomes higher at the inner wall at higher S. Going back to Fig. 6 again for S=0.82, the flame appears to be stable at t=9.0s, and the positive pressure gradient (caused due to vortex breakdown) along the inner wall pushes the flame tip into the mixing tube and starts interacting with incoming cross flow fuel injection (Fig. 6, t=9.0s-t=12.2s). Once the flame front starts moving upstream, due to complex interaction it starts forming low velocity recirculation region ahead of flame tip (Fig. 6), giving rise to the production of negative axial velocity to favor this upstream propagation severely. Whereas in case of S=0.38, the inner wall pressure gradient becomes negative and pushes the flame front back at later time instants (Fig. 5, t=14.2s, 17.6s), never leads to such upstream movement.

At S=1.76 (Fig. 7) compared to S=0.82 case (Fig. 6), the flame is already moved deep into upstream mixing tube at the earlier time instant of t=7.6s (Fig. 7) including distinct formation of two recirculation bubbles due to higher pressure gradient at the inner wall. One recirculation bubble, formed due to vortex breakdown, pushes the flame front upstream and then as the flame front moves further upstream it starts interacting with the fuel injection system and forms another recirculation bubble ahead of flame tip. This gives rise to the production of higher induced negative velocity that promotes upstream flame propagation aggressively at higher swirl strength (S=1.76). At t=9.2s in Fig. 7, the flame front already reaches the fuel-injection system. The observation in Figs. 5-7 confirms that the upstream flame propagation becomes favorable with the increase in swirl strength.

To understand the effects of heat release on the flow field, the vorticity transport equation (Eq. 26) which essentially shows the evolution of the vorticity of a moving fluid element in space is explored and written as:

$$\frac{D\vec{\omega}}{Dt} = \underbrace{(\vec{\omega} \cdot \vec{\nabla})\vec{u}}_{I} - \underbrace{\vec{\omega}(\vec{\nabla} \cdot \vec{u})}_{II} - \underbrace{\frac{\vec{\nabla}p \times \vec{\nabla}\rho}{\rho^2}}_{III} + \underbrace{\nu\nabla^2\vec{\omega}}_{IV} \tag{26}$$

where the RHS terms are: (I) Vortex stretching, (II) Gas expansion, (III) Baroclinic production, and (IV) Viscous diffusion. The terms (I) and (IV) have influence regardless of reacting or non-reacting flows. The term (IV) sharply rises across the flame due to change in temperature, and thus enhances the rate of diffusion and dampens the vorticity. However, the misalignment of the pressure and density gradients due to inclination and expansion of the flame with respect to the flow field contributes to the baroclinic production (III) of vorticity. The gas expansion term (II), acts as a sink in reacting cases, is directly proportional to the gas dilatation ratio across the flame ($\rho_u/\rho_b$) and increases as the temperature increases in presence of combustion. Hence two terms (II & IV) stabilize each other influences (Fig. 8) in reacting flow field. Therefore, the production of negative azimuthal vorticity (Fig. 8) at the inner edge of the flame (burnt region along the pipe wall) and also along the flame surface, is primarily due to interaction between shear generated vorticity and flame generated, baroclinic vorticity. While for S=0.38 (Fig. 8(a)), sufficient negative azimuthal vorticity is not produced along the flame front which can help the upstream flame propagation and makes this case stable without showing any upstream flame movement. In fact for this case, the baroclinic production gives rise to positive azimuthal vorticity and pushes back the flame front by generating positive induced velocity, whereas in the case of S=0.82 (Fig. 8(b)) baroclinic production is accompanied by the stretching term and both these terms give rise to the production of sufficient negative azimuthal vorticity that promotes upstream flame propagation. For S=1.76 (Fig. 8(c)) compared to S=0.82, this effect is very dominant and aggressively favors the upstream flame movement.

Moreover, it is worthwhile to mention that the negative azimuthal vorticity induces a negative axial velocity in the flow filed following the relationship as given

$$w_{ind}(x) = \frac{1}{2} \int_{-\infty}^{\infty} \int_{0}^{\infty} \frac{r'^2 \, \omega_\theta(r', x')}{\left[ r'^2 + (x - x')^2 \right]^{3/2}} dr' dx' \qquad (27)$$

Hence this clearly states that the greater the vorticity, the greater is the induced velocity. Thus, the higher azimuthal vorticity produces higher negative induced velocity which pushes the stagnation point ahead of the flame tip (Figs. 6-7, for S=0.82 & 1.76) further upstream and helps to form a small recirculation bubble primarily due to pressure jump across the convex flame orientation in the flow field The formation of this small recirculation bubble ahead of the flame tip gives rise to the generation of greater negative azimuthal vorticity, which in turn, produces higher induced negative velocity and the flame tip movement to further upstream becomes completely uncontrollable at higher swirl strength, i.e. S=0.82 and S=1.76.

A more detailed observation of vorticity budget terms (Eq. 26) supports the above phenomena, which contributes to the generation of negative azimuthal vorticity, in turn produces the induced negative velocity. Figure 8(b-c) shows the distributions of change in budget terms for S=0.82 (between t=9.0s and t=12.2s) and S=1.76 (between t=7.6s and t=9.2s). The upstream propagation of the flame front is due to increase of the induced negative velocity (Eq. 27). A thorough analysis clearly exhibits that the combined effects of vortex stretching and baroclinic production give rise to the negative azimuthal vorticity particularly along the flame front, while the vortex expansion and diffusion terms stabilize each other influences. Since negative azimuthal vorticity induces negative axial velocity, vortex stretching and baroclinic production are primarily responsible for upstream flame propagation at higher swirl strength (S=0.82 & S=1.76). Moreover, Figs. 6-7 also support this fact that the flame tip encounters much higher negative induced velocity in between t=12.2s & t=9.0s for S=0.82 and t=7.6s & t=9.2s for S=1.76 due to combined effect of vortex-stretching and baroclinic production and that increases consistently. Whereas for S=0.38 (Fig. 8(a)), baroclinic production contributes to the production of positive azimuthal vorticity along the flame front and that's why flame front is pushed backed as shown in Fig. 8, and produces stable movement of flame front for this case.

### 4.2.2   Effects of premixedness

The flow configuration used for this parametric study has fixed swirl strength of S=0.82. Once set-up has jet-in-cross flow fuel injection system as discussed in the previous subsection; means the fuel is injected right after the step (Fig. 1) and the inlet swirled air is mixed with the injected fuel to have a premixed mixture at dump plane. Other set-up assumes everything is premixed at the upstream inlet (Fig. 1) based on φ=0.7.

Figure 9 shows the time series temperature plots for both perfectly premixed inlet and jet-in-cross flow fuel injection systems at S=0.82.  As observed in both the systems, the flame moves upstream of the dump plane to premixing section and stabilizes there. Notably, the setup with perfectly premixed inlet appears to be more susceptible to this behavior. This is associated with the premixedness of the fuel-air mixture as well as swirl strength that enhances turbulence-chemistry interaction, producing higher heat-release in hydrogen enriched mixture and the upstream flame movement is the consequence. In the case of perfectly premixed inlet, the flame front has moved into the mixing tube upto X/2D~-2 at very earlier instant (t=9.0s) due to recirculation bubble movement, while in the other case (jet-in-cross flow) the flame front has only moved upto X/2D~-0.5 at same time instant. For perfectly premixed case rich fuel-air mixture in addition to higher turbulence levels (S=0.82) induces higher flame propagation velocity compared to other case, and exhibits to be more susceptible to this behavior. In addition, the vortex breakdown is often accompanied by a large recirculation zone with high reverse flow velocities in combustions systems, and the reverse velocities can promote upstream flame propagation (Fig. 9).

As noted and discussed earlier, the formation of recirculation zone is the consequence of vortex breakdown. The existence of this recirculation bubbles ahead of flame tip provides enough ingredient to favor upstream flame propagation. The initial flame front movement to the mixing tube is caused due to unsteady motion of recirculation bubble (due pressure jump across the flame front), while the later movement is accompanied primarily due to induced negative velocity. Fig. 10(a) shows the time instant snap shots of stream lines superimposed with temperature contours. . Compared to jet-in-cross flow case (Fig. 9), the flame is already moved deep into upstream mixing tube at the time instant of t=9.0s (Fig. 10(a)) including upstream stagnation point along the bluff-body wall. Evidently vortex break down due to chemical reaction (CIVB) is more intensive in this case and thus making this system more susceptible to such behavior. Due to perfectly premixed inlet, the flame tip encounters much more fuel rich condition compared to jet-in-cross flow configuration, thereby enhances chemical reaction. Also, the presence of highly combustible $H_2$ in fuel mixture makes the upstream flame movement much more severe in this case compared to other set up (jet-in-cross flow).

To further illustrate the above hypothesis, it is worthwhile to look at the distribution of vorticity budget terms (Eq. 26) which clarifies the upstream flame tip movement at later time instances. The net production of negative azimuthal vorticity produces the induced negative velocity. Fig. 10(b) shows the time difference distribution of budget terms between t=9.0s and t=15.0s for perfectly premixed case. This supports the fact that the flame tip encounters much higher negative induced velocity in between t=9.0s and t=15.0s due to combined effect of vortex-stretching and baroclinic production and that increases consistently. This upstream propagation of the flame front is due to increase of the induced negative velocity (Eq. 27). The analysis clearly reveals that the combined effects of vortex stretching and baroclinic production give rise to the negative azimuthal vorticity particularly along the flame front, while the vortex expansion and diffusion terms stabilize each other influences. Since negative azimuthal vorticity induces negative axial velocity, vortex stretching and baroclinic production are primarily responsible for upstream flame propagation and leading to upstream flame movement for this case. However, jet-in-cross flow case also exhibits similar behavior but the upstream flame propagation is only delayed compared to perfectly premixed case.

### 4.2.3   Effects of geometry (step )

In order to study the effect of geometry, the flow configuration considered here with perfectly premixed inlet (assumes everything is premixed at the upstream inlet (Fig. 1) based on φ=0.7) and having fixed swirl strength of S=0.82. Once set-up has the 'step' in upstream mixing tube, other one does not have any 'step' in the mixing tube.

Figure 11(a) shows the time series temperature plots for the case 'without step'. Comparing with the case with 'step' (Fig. 9), it is evident that both the systems exhibit upstream flame movement, while the setup with 'step' rather shows more susceptible to such behavior. The interesting phenomenon observed here is that the temperature in the product region (behind the center-body) is less for the case ('without step') due to higher flow acceleration in the upstream delivery tube in presence of lower annulus (without the step in the geometry), whereas the other setup ('step') produces much higher temperature regions due to existence of low velocity in B/L in presence of step. Although, both the system shows similar flame behavior, the burner 'without step' may be useful for lowering NOx production due to lower temperature generation in the burnt regions.

The key thing is to be noted here is that the upstream flame propagation for this case ('without step') is delayed compared to other case ('step', Fig. 9). As noticed for this case ('without step'), the flame front has only reached upto X/2D~-0.5 into the mixing tube at t=9.0s due to recirculation bubble movement, while in the other case ('step') the flame front has moved upto X/2D~-2.0 at same time instant (Fig. 9). These observations confirm the fact that the step has an impact on upstream flame movement. Since both the cases have perfectly premixed mixture, the flame front always encounters rich fuel region during its upstream movement, while the presence of 'step' (Fig. 9) provide some additional favorable condition by generating low velocity region in the B/L along the bluff-body wall. Evidently for this case ('without step') also, upstream flame propagation occurs due to vortex breakdown, which is less intensive compared to other case ('step') as shown in Fig. 9. The reason behind this is the negative induced velocity becomes much higher than the low axial velocity in B/L, formed due to 'step'; and thereby aggressively favors the upstream flame movement for the setup with 'step' (Fig. 9) whereas in the other case ('without step') incoming flow accelerates more (lower annular area) and thereby avoiding such situation resulting delayed occurrence of upstream movement (Fig. 11(a)). Similar analysis of vorticity budget terms (Eq. 26) supports the above phenomena, which contributes to the generation of negative azimuthal vorticity, in turn produces the induced negative velocity. Fig. 11(b) depicts the time difference distribution of budget terms between t=10.2s and t=15.2s for the case 'without step'. Expectedly, the analysis reveals the fact that the combined effects of vortex stretching and baroclinic production give rise to the negative azimuthal vorticity particularly along the flame front, while the vortex expansion and diffusion terms stabilize each other influences. Here also, it can be said that vortex stretching and baroclinic production are primarily responsible for upstream flame propagation and leading to flashback. Comparing both the cases (step & without step), one should note that both these configurations behave in a similar fashion but the upstream movement takes place at different time instance due to mixing tube flow dynamics. Hence, the observed phenomena confirm that change in geometry definitely has some impact on CIVB, but not good enough to avoid occurrence of upstream flame propagation with hydrogen enriched mixture.

### 4.2.4    Effects of geometry (step ) and swirl

To complete the loop of parametric study, it is worth looking at the effects of different swirl strengths S=0.38 and S=0.82 with perfectly premixed inlet and without a step geometry.

The instantaneous temperature contours for different swirl strength with perfectly premixed inlet is shown in Figure 12. Expectedly, the setup with higher swirl (S=0.82) exhibits upstream flame movement as discussed in the previous subsection. Moreover, the effect of different swirl strength with jet-in-cross flow fuel injection is also discussed in the subsection 'effect of swirl' where the key observation was that the flame does not propagate upstream at lower swirl strength (S=0.38). Here in the present scenario, the setup is perfectly premixed (fuel rich condition compared to jet-in-cross flow fuel injection) and operated under lower swirl strength (S=0.38). The result reveals the same fact that the flame does not exhibit any upstream movement at lower swirl strength (S=0.38) even with perfectly mixture. Hence, it can be confirmed that systems with higher swirl become more susceptible to upstream flame propagation, especially with hydrogen enriched mixtures. Fig. 12 supports the impact of swirl on instantaneous flame propagation compared to Fig. 11 (a). As observed in Fig. 12, the flame front is never pushed into the mixing tube due to CIVB (as discussed earlier subsections). Only the flame tip shows a little upstream movement (Fig. 12), but thereafter it does not encounter any favorable conditions to promote further upstream movement; such as turbulence, low velocity B/L. Lowering swirl intensity implicitly means lowering the level of turbulence, and thus altering the chemical reaction to reduce the effects of CIVB. Obviously, in the case with low swirl, azimuthal vorticity production is lower (as discussed under the subsection 4.2.1) and thus does not favors any upstream flame movement. Even though highly combustible hydrogen enriched mixture does not exhibit such behavior.

In summary, the above observations confirm the parametric effects on CIVB and flame behavior for hydrogen enriched prefixed flame. The flame contributes to vortex breakdown, and results a low or negative flow region ahead of it (recirculation bubble formation with a stagnation point). As the flame tip moves forward, causing the vortex location breakdown region to advance further upstream. This process continues as the flame proceeds further and further upstream. In majority, swirl has a huge impact on CIVB as well as premixedness upto certain extent, especially for hydrogen enriched premixed flames.

## 5. CONCLUSIONS

LES with a TF model is used to investigate hydrogen enriched premixed flames in a laboratory based model combustor. The effect of swirl, geometry and premixedness is investigated for hydrogen enriched premixed flame, especially in the context upstream flame movement. The upstream flame propagation behavior for swirled burner is investigated and reported here on the basis of the of the source terms of the vorticity transport equations. A 2-step chemical scheme for methane combustion and 1-step for hydrogen combustion are invoked to represent the flame chemistry for methane-hydrogen-air combustion. The equivalence ratio for the flame is 0.7 and the Reynolds number is Re=13339 with 30%$H_2$ mixture. This study leads to the following conclusions:

(a) LES-TF model is able to properly capture hydrogen enriched combustion behavior.

(b) The reacting velocity profiles are well predicted.

(c) Higher swirl broadens the size of recirculation zone for a fixed $H_2$ enriched mixture.

(d) Increase in swirl strength always leads to upstream flame propagation due to higher turbulence, thus enhances chemical reaction.

(e) Premixedness upto certain extent also affects such behavior. At a particular swirl strength (S=0.82), perfectly premixed mixture tends to behave more susceptible to such behavior.

(f) Geometry change has a little impact on flame behavior at S=0.82.

(g) The vortex stretching and baroclinic production contribute to the net vorticity generation which produces considerable levels of negative axial velocity that favors upstream flame propagation.

(h) If the flame does not produce enough baroclinic torque, then the combined effects of gas-expansion and diffusion can stabilize the vortex flow and prevent the upstream movement of the flame. At lower swirl strength the baroclinic torque is balanced by vortex stretching and does not produce any favorable condition for such behavior.

(i) The present study reveals that swirl and premixedness has effect on upstream flame propagation behavior in a center-body stabilized swirled burner.

This study demonstrates that the Thickened-Flame based LES approach with simplified chemistry for reacting flows is a promising tool to investigate reacting flows in complex geometries.

## ACKNOWLEDGMENTS

The authors would like to thank Shengrong Zhu for proving the experimental data. This work was supported by the Clean Power and Energy Research Consortium (CPERC) of Louisiana through a grant from the Louisiana Board of Regents. Simulations are carried out on the computers provided by LONI network at Louisiana, USA (www.loni.org) and HPC resources at LSU, USA (www.hpc.lsu.edu). Finally, the manuscript preparation as well as partial analysis of data is carried out using the resources available at Indian Institute of Technology Kanpur (IITK), India. This support is gratefully acknowledged.

## Nomenclature

A       pre-exponential constant

$C_s$      LES model coefficient

$D_i$      molecular diffusivity

E       efficiency function

$E_a$      activation energy

$S_{ij}$      mean strain rate tensor

$T_a$      activation temperature

U       mean axial velocity

$U_o$      bulk inlet velocity

| $u_i$ | velocity vector |
| $u'$ | rms turbulence velocity |
| W | mean tangential velocity |
| w' | tangential RMS velocity |
| $x_i$ | Cartesian coordinate vector |
| $Y_i$ | species mass fraction |

**Greek symbols**

| $\Delta$ | mesh spacing |
| $\nu_t$ | kinematic turbulent eddy viscosity |
| $\bar{\rho}$ | mean density |
| $\omega_i$ | reaction rate |

**List of Figure captions**

Figure 1. Schematic view of computational domain for the swirl injector

Figure 2. Non-reacting flow results for Re=13339 at different axial locations [D=Center-body diameter]: Experimental data ($\Delta$), Lines are LES predictions: fine mesh ( —— ), coarse mesh ( - - - - ). Mean axial velocity $U/U_o$, Mean tangential velocity $W/U_o$, Axial velocity fluctuation $u_{rms}/U_o$, Tangential velocity fluctuation $w_{rms}/U_o$

Figure 3. Reacting flow results: (a) Recirculation bubble size for different S, (b) Velocity profiles for S=0.82 at different axial locations: Experimental data ($\Delta$), Lines are LES predictions: Mean axial velocity $U/U_o$, Axial velocity fluctuation $u_{rms}/U_o$

Figure 4. Instantaneous flame propagation for different S (Temperature in K scale)

Figure 5. Stream lines colored with temperature contours (K scale) for S=0.38: no upstream flame movement: isotherms in black lines (300, 675, 1050, 1425 K)

Figure 6. Stream lines colored with temperature contours (K scale) for S=0.82 during upstream flame movement: isotherms in black lines

Figure 7. Stream lines colored with temperature contours (K scale) for S=1.76 during upstream flame movement: isotherms shown black lines

Figure 8. Individual terms of vorticity transport equation (Eq. 26) along the flame arc length of isotherm (1050K) for different S [0=center-body wall (flame tip)]: (a) S=0.38 (no upstream flame movement), (b) S=0.82 during upstream flame movement, (c) S=1.76 during upstream flame movement

Figure 9. Instantaneous flame propagation for S=0.82: (a) perfectly premixed inlet, (b) jet-in-cross-flow fuel injection (Temperature in K scale)

Figure 10. Perfectly premixed inlet with S=0.82 during upstream flame movement: (a) Stream lines colored with temperature contours (K scale, isotherms in black lines), (b) Individual terms of vorticity transport equation (Eq. 26) along the flame arc length of isotherm (1050K) [0=center-body wall (flame tip)]

Figure 11. Perfectly premixed inlet with S=0.82 & without-step during upstream flame movement: (a) Instantaneous flame propagation (Temperature in K scale), (b) Individual terms of vorticity transport equation (Eq. 26) along the flame arc length of isotherm (1050K) [0=center-body wall (flame tip)]

Figure 12. Instantaneous flame propagation for S=0.38: perfectly premixed inlet & without-step (Temperature in K scale)

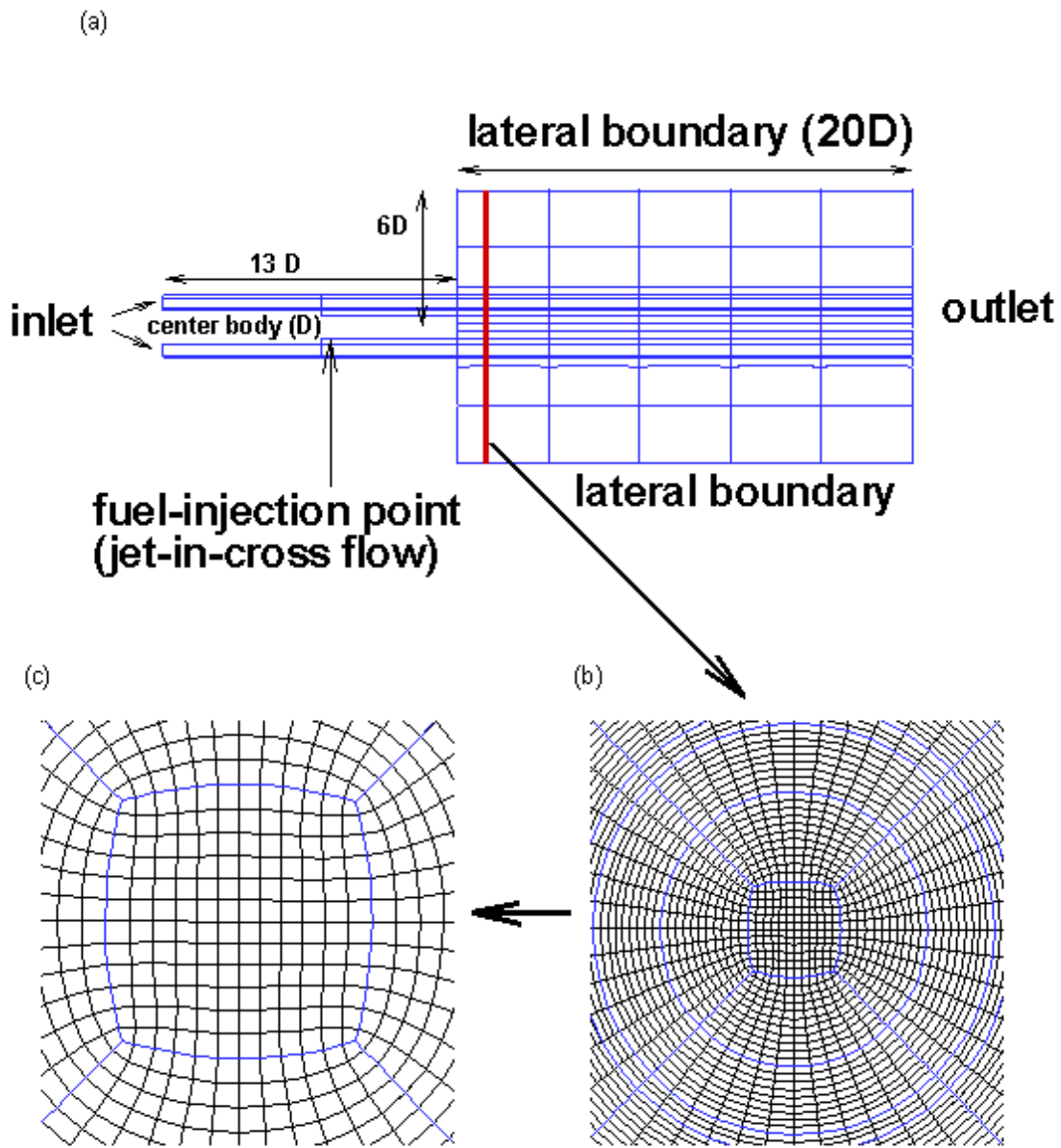

Figure 1

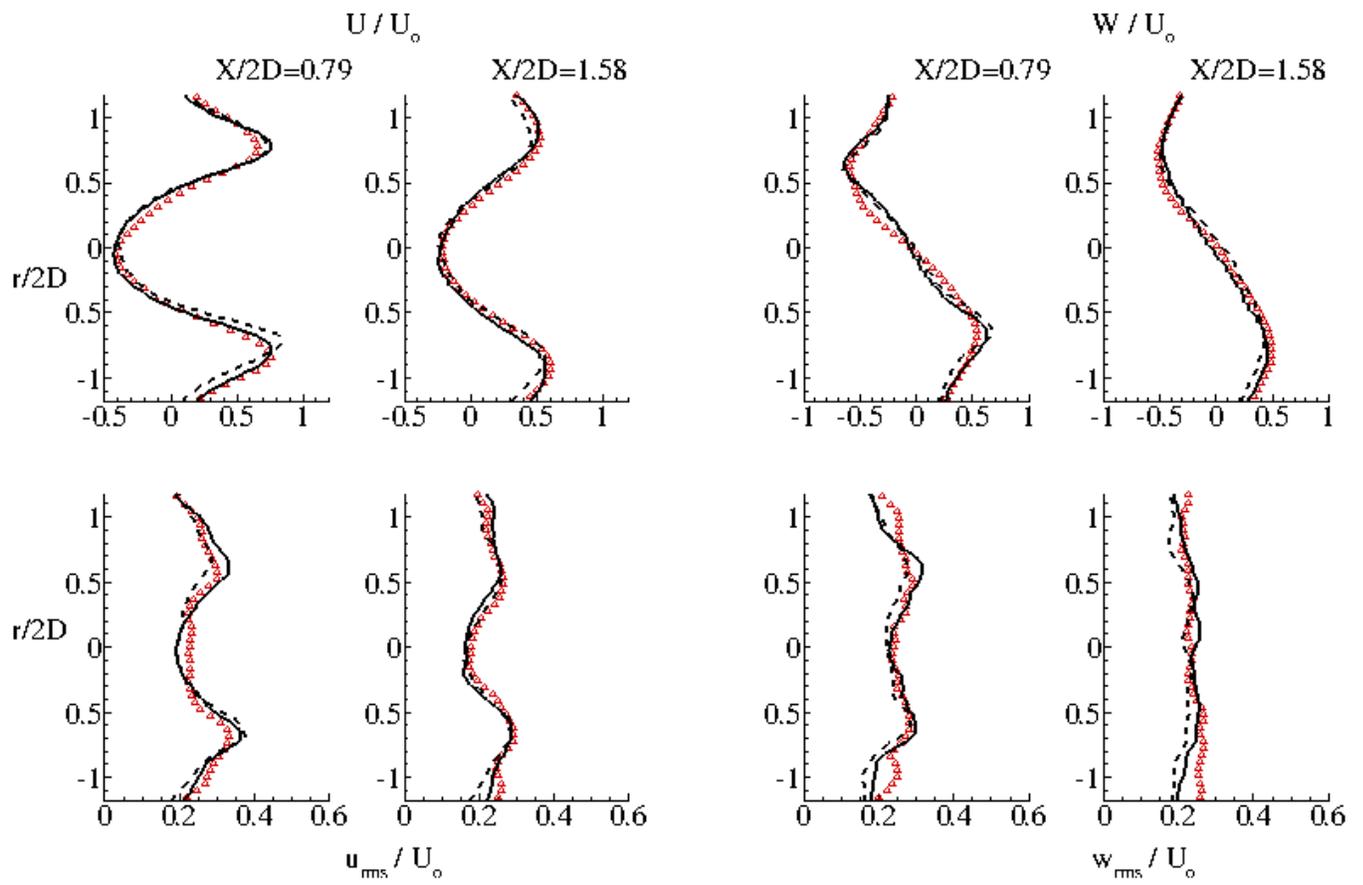

Figure 2

(a)

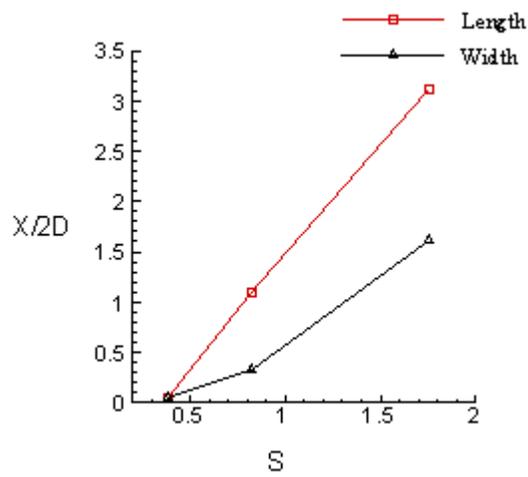

(b)

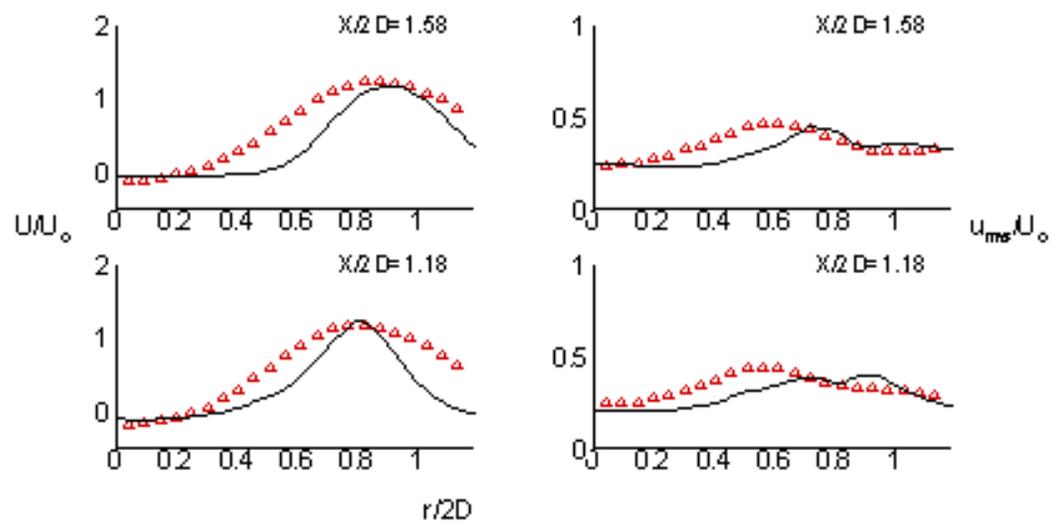

Figure 3

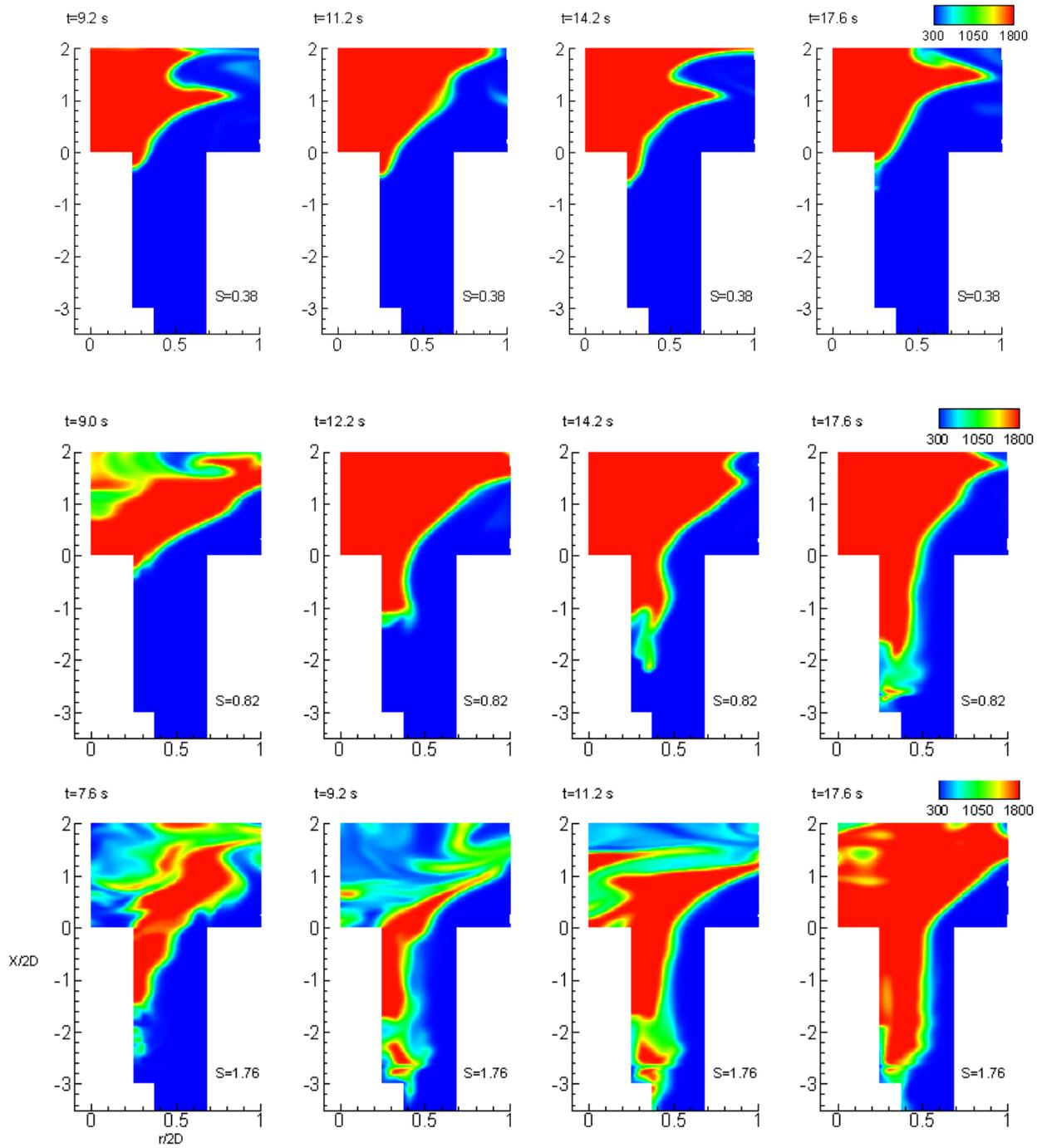

Figure 4

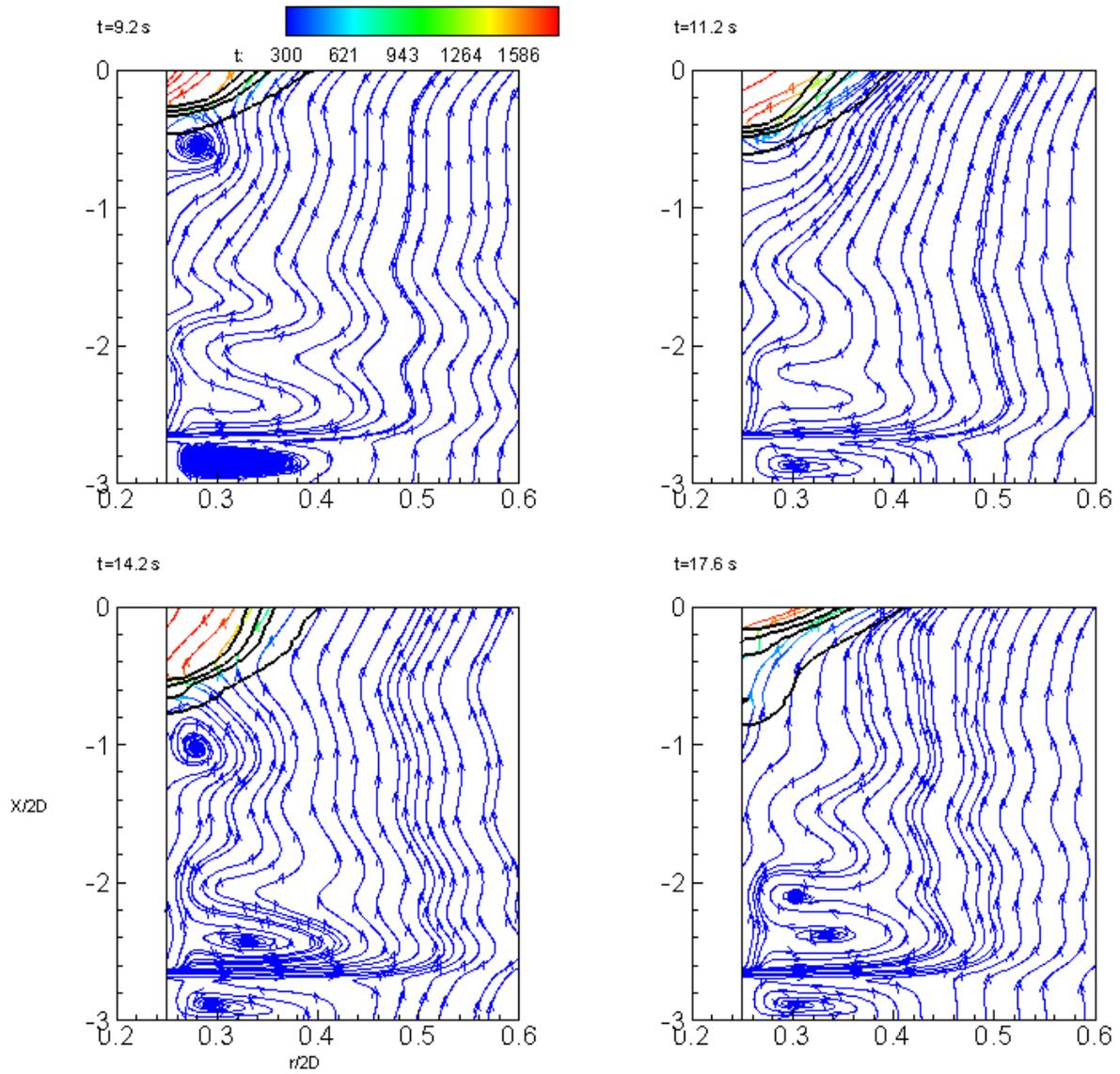

Figure 5

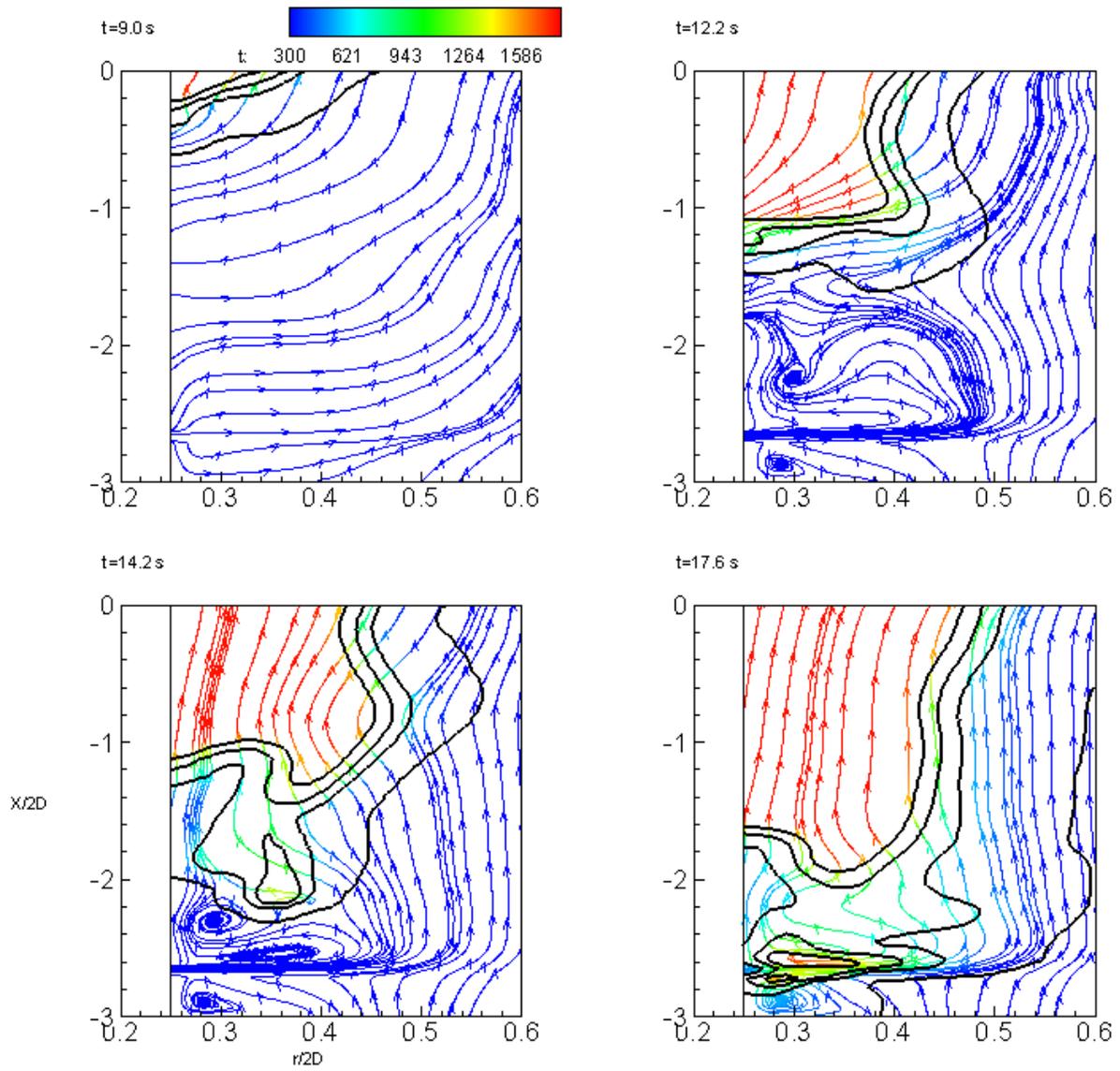

Figure 6

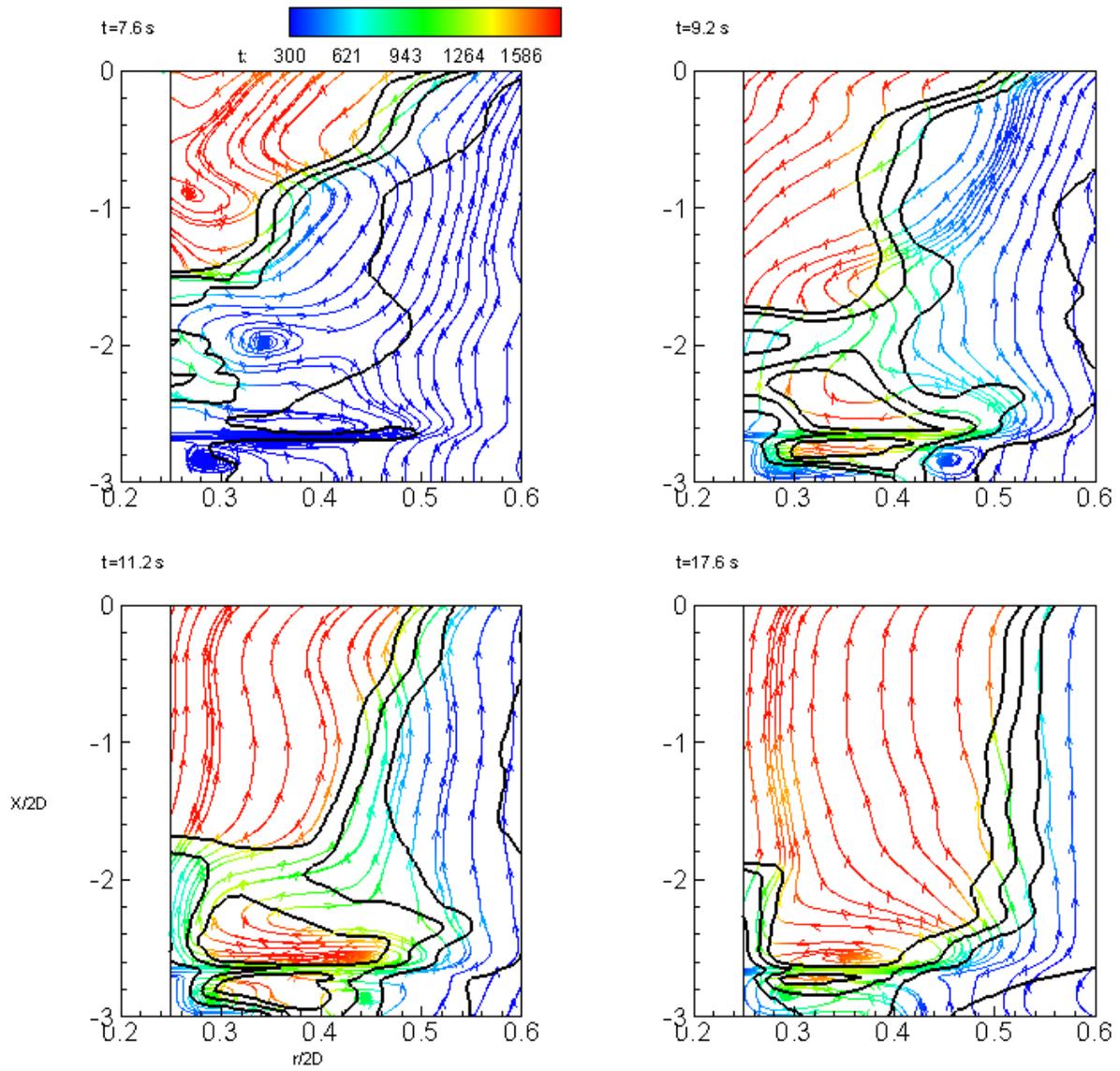

Figure 7

(a)

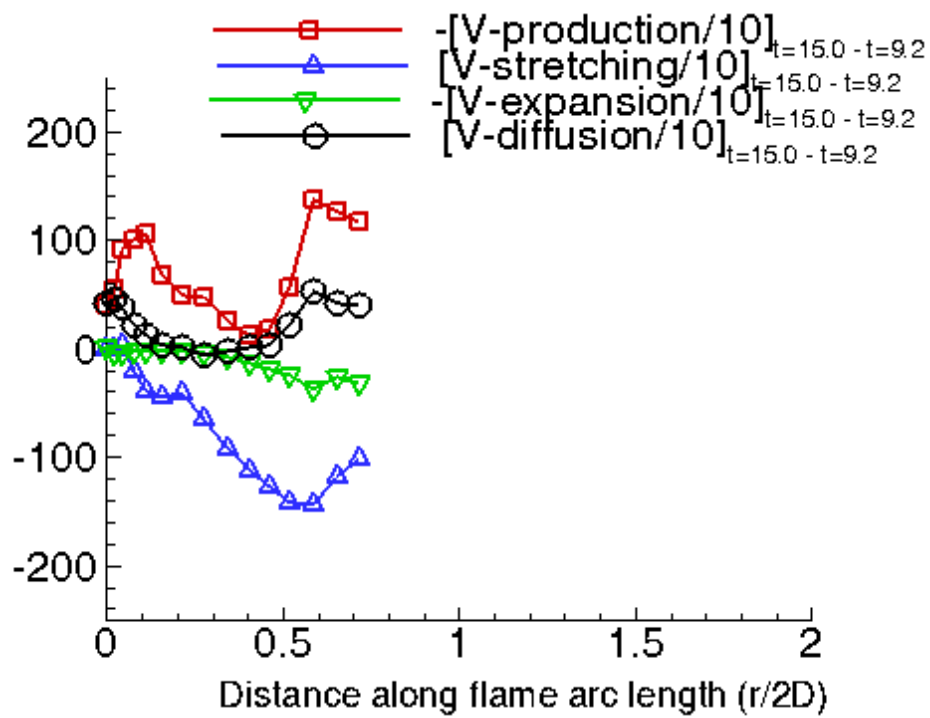

(b)

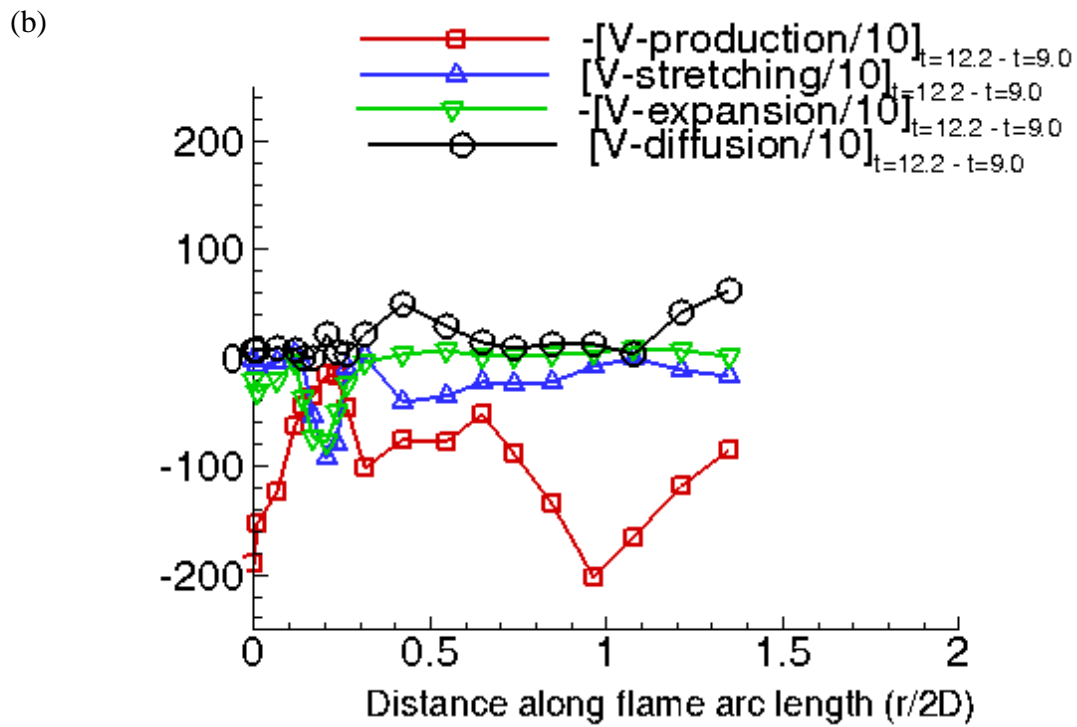

(c)

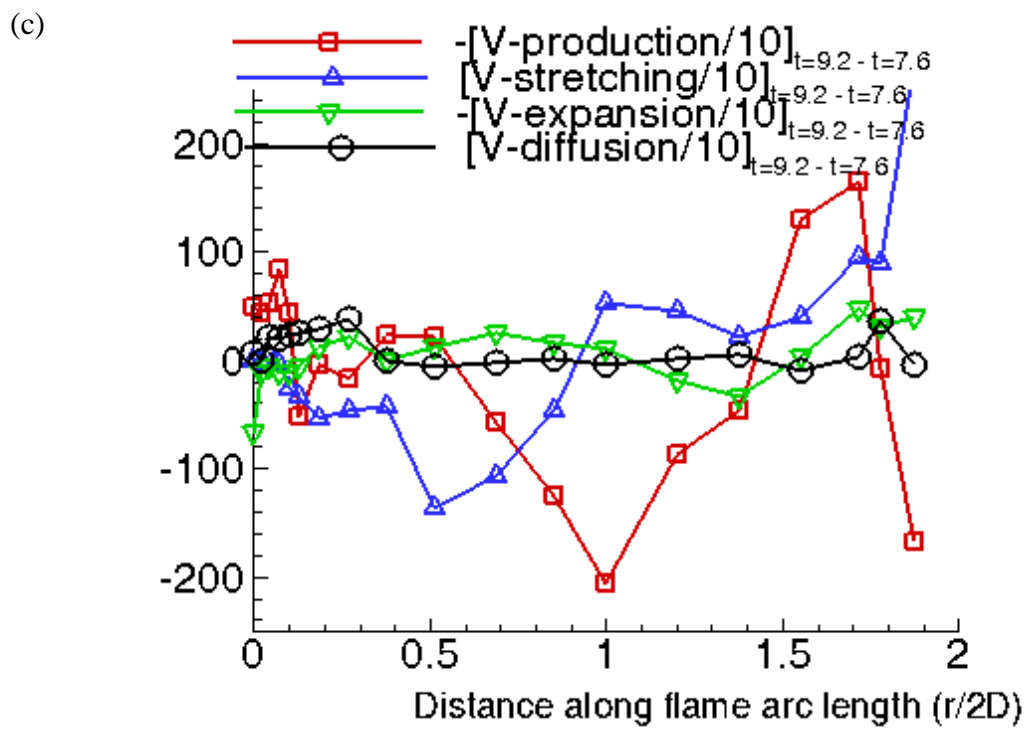

Figure 8

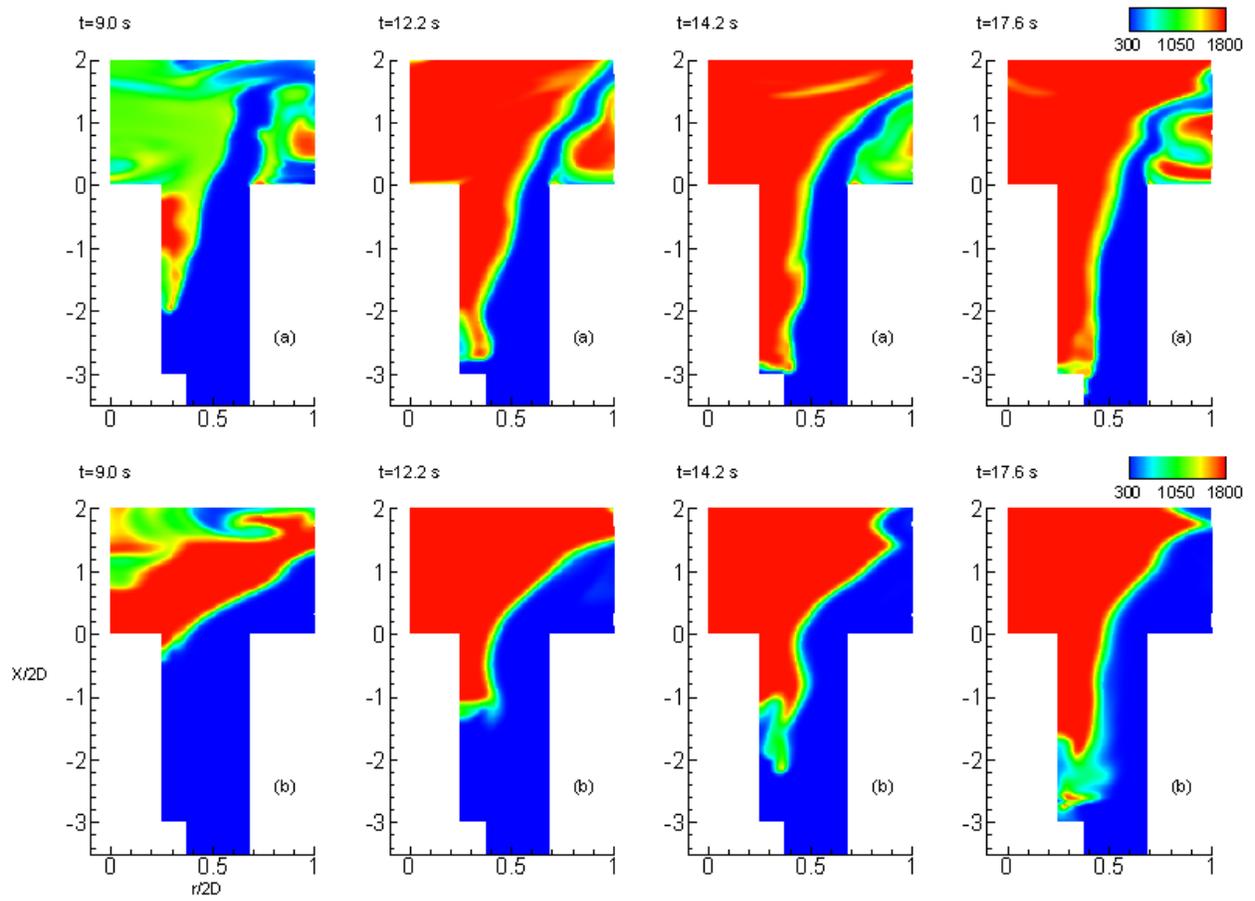

Figure 9

(a)

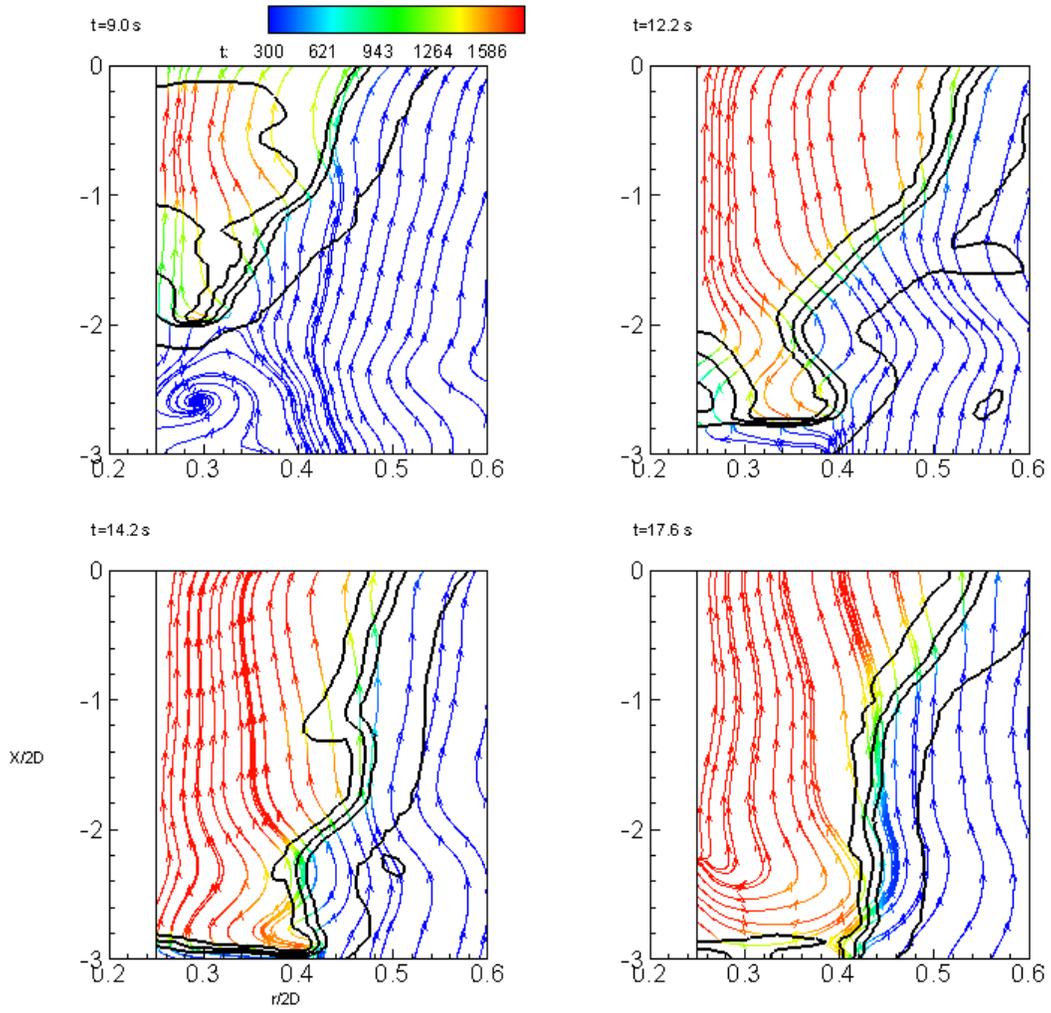

(b)

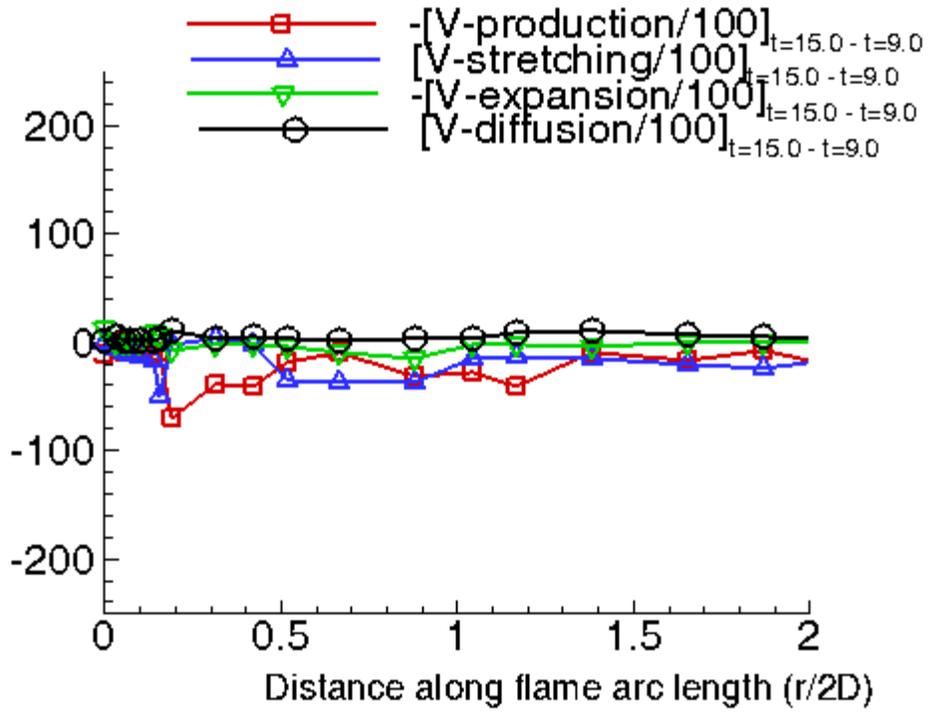

Figure 10

(a)

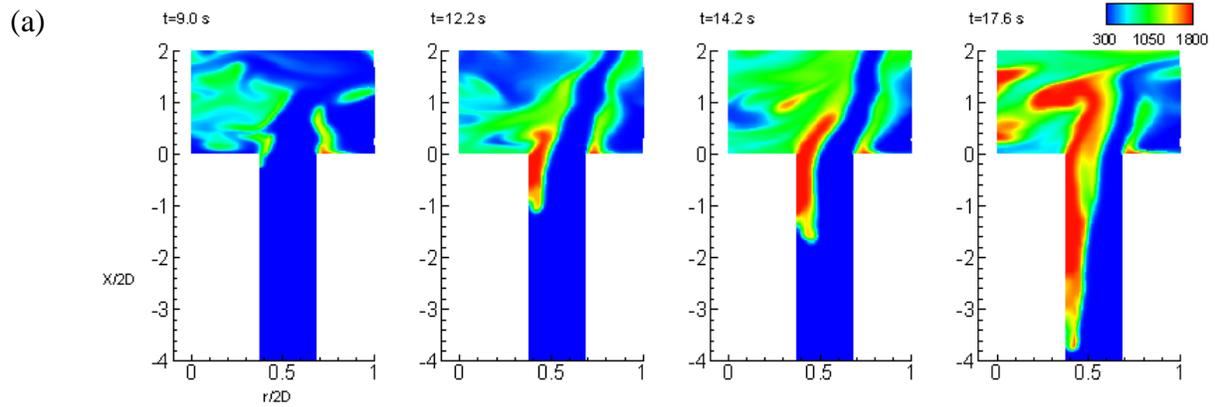

(b)

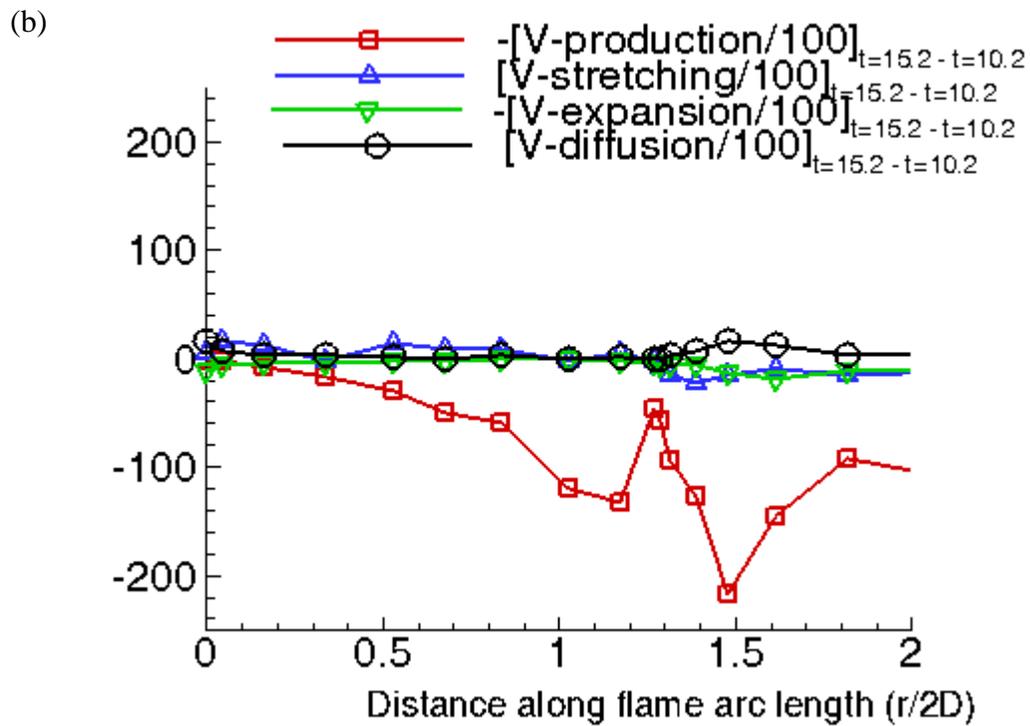

Figure 11

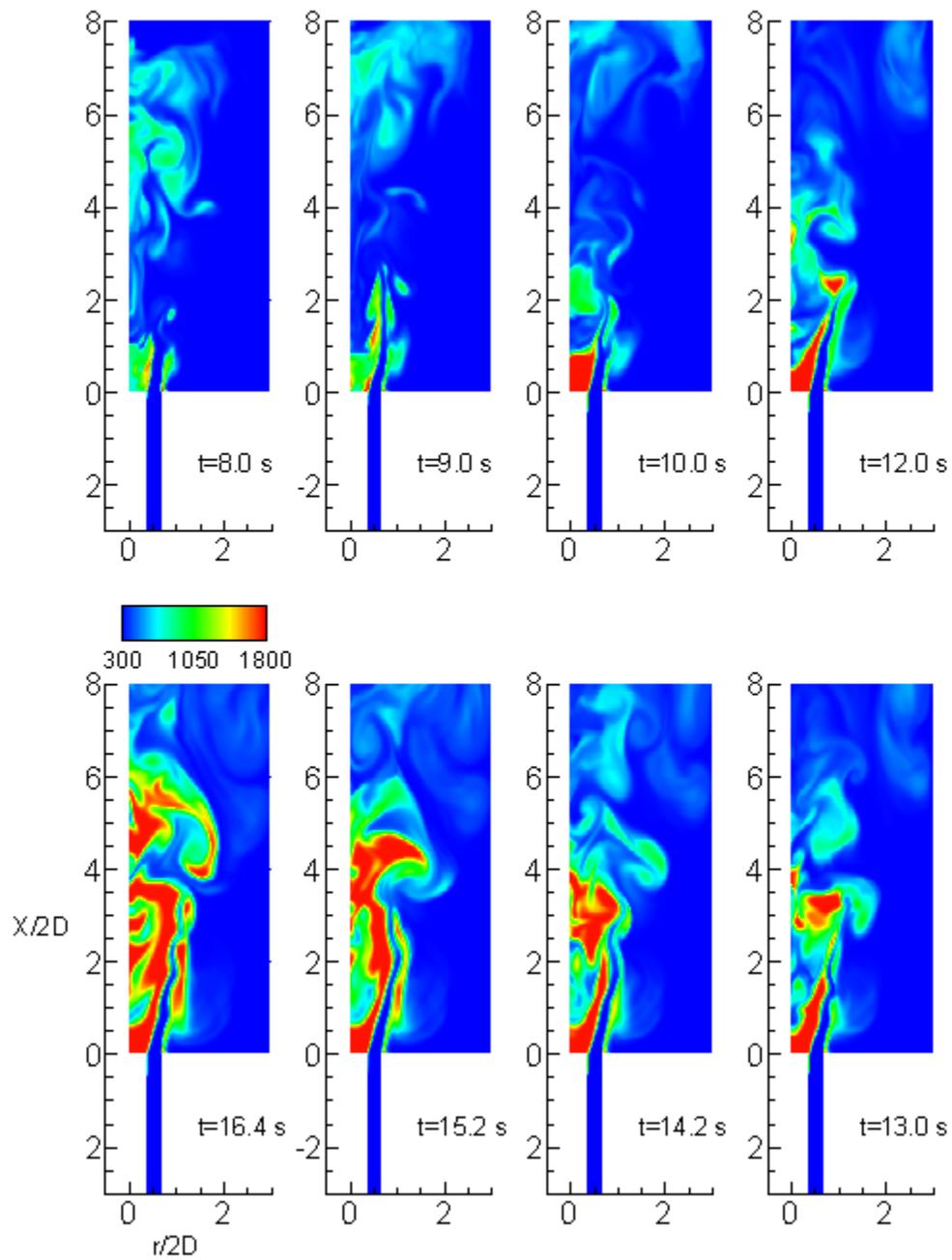

Figure 12